\documentclass[10pt,journal,compsoc]{IEEEtran}
\IEEEoverridecommandlockouts
\usepackage{cite}
\usepackage{amsmath,amssymb,amsfonts}
\usepackage{amsthm}
\usepackage{algorithmic}
\usepackage{graphicx}
\usepackage{textcomp}
\usepackage{xcolor}
\usepackage{caption}
\usepackage{multirow}
\usepackage{hhline,subfigure}
\usepackage[utf8]{inputenc}
\usepackage{algorithm}
\def\BibTeX{{\rm B\kern-.05em{\sc i\kern-.025em b}\kern-.08em
    T\kern-.1667em\lower.7ex\hbox{E}\kern-.125emX}}
    
\newtheorem{theorem}{Theorem}[section]

\newtheorem{definition}{Definition}[section]

\begin{document}

\title{
Incentive Mechanism Design for Joint Resource Allocation in Blockchain-based Federated Learning
\author{Zhilin Wang, Qin~Hu, Ruinian Li, Minghui Xu, and Zehui Xiong}
\thanks{This work is partly supported by the US NSF under grant CNS-2105004.}
\IEEEcompsocitemizethanks{
\IEEEcompsocthanksitem Zhilin Wang and Qin Hu (corresponding author) are with the Department of Computer and Information Science, Indiana University-Purdue University Indianapolis, IN, 46202, USA. E-mail: \{wangzhil,qinhu\}@iu.edu

\IEEEcompsocthanksitem Ruinian Li is with the Department of Computer Science, Bowling Green State University, Bowling Green, Ohio, 43551, USA. E-mail: lir@bgsu.edu

\IEEEcompsocthanksitem Minghui Xu is with the School of Computer Science and Technology, Shandong University, China. E-mail: mhxu@sdu.edu.cn

\IEEEcompsocthanksitem Zehui Xiong is with Pillar of Information Systems Technology and Design,
Singapore University of Technology Design, Singapore. E-mail: zehui\_xiong@sutd.edu.sg
}
}






\IEEEpubidadjcol

\IEEEtitleabstractindextext{

\begin{abstract}

Blockchain-based federated learning (BCFL) has recently gained tremendous attention because of its advantages such as decentralization and privacy protection of raw data. However, there has been few research focusing on the allocation of resources for clients in BCFL. In the BCFL framework where the FL clients and the blockchain miners are the same devices, clients broadcast the trained model updates to the blockchain network and then perform mining to generate new blocks. Since each client has a limited amount of computing resources, the problem of allocating computing resources into training and mining needs to be carefully addressed. In this paper, we design an incentive mechanism to assign each client appropriate rewards for training and mining, and then the client will determine the amount of computing power to allocate for each subtask based on these rewards using the two-stage Stackelberg game. After analyzing the utilities of the model owner (MO) (i.e., the BCFL task publisher) and clients, we transform the game model into two optimization problems, which are sequentially solved to derive the optimal strategies for both the MO and clients.
Further, considering the fact that local training related information of each client may not be known by others, we extend the game model with analytical solutions to the incomplete information scenario. Extensive experimental results demonstrate the validity of our proposed schemes.
\end{abstract}
}

\maketitle

\section{Introduction}
\IEEEPARstart{S}INCE its emergence in 2016, federated learning (FL) has been greatly developed and widely applied in many fields, such as Internet of Things \cite{zhao2020mobile,zhang2020blockchain,lu2019blockchain}, smart transportation \cite{qi2021privacy,hua2020blockchain} and healthcare \cite{passerat2019blockchain,aich2021protecting,kumar2021blockchain}. One of the most important advantages of FL is that there is no transmission of raw data from mobile devices to the centralized server for model training; instead, by training models on local devices (a.k.a., clients) and averaging all local models, FL significantly reduces the possibility of leaking data privacy to a large extent \cite{mcmahan2017communication}. However, there are still some challenges that may restrain the implementation and wide application of FL, e.g., the risk of the single point of failure, malicious attacks from participated clients, and the lack of participation incentives \cite{bagdasaryan2020backdoor, sattler2019robust,lyu2020threats,mothukuri2021survey}. 

In recent years, researchers resort to blockchain technology to tackle the challenges of FL, where  the blockchain system usually works as a decentralized system to provide incentives and data verification \cite{ramanan2020baffle, kim2019blockchain, liu2020fedcoin}. The combination of blockchain and FL is termed as blockchain-based FL (BCFL). In the BCFL framework, model updates submitted by clients will be verified by miners 
before the global aggregation algorithm is conducted. Once the global model is obtained, it will be updated into the main chain that can be accessed by all qualified participants. Though BCFL can partially address the aforementioned challenges of traditional FL, there are still some remaining issues that need to be addressed. 

One of the most critical problems in BCFL is  the resource allocation on local devices. Firstly, local devices with heterogeneous computational power usually have their own tasks to finish, so a universal resource allocation scheme for all the mobile devices is not practical. In addition, the whole system may not work effectively and sustainably if there are no reasonable rewards allocated to clients. Furthermore, both training and mining in the framework of BCFL consume significant amount of resources and time,  and thus it is difficult for clients to appropriately allocate their limited resources to ensure the performance of the global model during the required time period. Lastly, since the system may not be aware of the amount of training data that each client owns, it can be challenging for the model owner (MO), i.e., the BCFL task publisher, to make proper decisions regarding the reward distribution. 

There exist very few studies that tackle the above challenges\cite{Li2021, Hieu2020}. The existing studies are based on two assumptions which are not practical: 1) all clients have identical computational power and data volume; and 2) the system knows all the information about clients. 
To fill the gap, we propose an incentive mechanism for joint resource allocation on mobile devices in BCFL that is applicable to incomplete information scenario.

For the first challenge regarding the unbalanced distribution of resources on mobile devices, we let the clients decide how much computational power they are willing to devote into the training and mining tasks by themselves. By this means, clients can flexibly allocate computation resources for their own tasks. In addition, in our model, training and mining are performed sequentially, and the amount of computational power devoted to these two tasks can be different. 

To overcome the second challenge of motivating clients to join BCFL, we design an incentive mechanism to reward clients. Training and mining are two different tasks that require different amount of computational power, and thus the rewards should also be different. To ensure a fair distribution of rewards to all clients, we employ the approach of Shapely Value (SV) \cite{qu2020privacy} to determine the contributions of clients in the training process, which will affect the constraints in their respective optimization problems.

To address the last two challenges, we build the Stackelberg game model under both the complete and incomplete information situations, which are solved separately but with different insights. Based on the derived optimal solutions, our system can make optimal decisions in different information conditions.

In summary, our contributions in this work can be summarized as below:
\begin{itemize}
    \item We model the BCFL resource allocation problem as a two-stage Stackelberg game to help the MO make decisions on assigning how many rewards to each client for training and mining and to assist clients in determining the corresponding amount of computational power to be devoted in each subtask, via maximizing their respective utilities.
    \item In order to maintain the stability and sustainability of the whole BCFL system, we design a fair reward allocation scheme inspired by SV to calculate the rewards for clients based on their contributions in the training process.
    \item Considering that the training related information of devices may not be known to others in the practical application scenario, we further study the resource allocation mechanism under the incomplete information situation and derive the optimal solutions accordingly.
    \item We test our proposed resource allocation mechanisms through extensive experiments. The experimental results show that these mechanisms are effective. 
\end{itemize}

The rest of this paper is organized as follows. We introduce the system model and problem formulation based on the two-stage Stackelberg game in Section \ref{sec:formulation}. The detailed models and solutions under complete and incomplete information scenarios are reported in Section \ref{sec:game1} and Section \ref{game2}, respectively.  Experimental evaluations are presented in Section \ref{sec:experiment}. We present the related work in Section \ref{sec:related}. Finally, we conclude this paper in Section \ref{sec:conclusion}.

\section{System Model and Problem Formulation}\label{sec:formulation}

In this section, we will illustrate the system model of our considered blockchain-based federated learning (BCFL) and then formulate the problem from the perspective of resource allocation and incentive mechanism design based on Stackelberg game. 
For convenience, we list the key notations in Table \ref{table:not}.

\begin{table}[h!]
\caption{Key Notations.}
\centering
\begin{tabular}{|c|l|}
\hline
Notation      & \multicolumn{1}{c|}{Meaning}                                                                                                               \\ \hline
$\mathcal{N}$ & The set of clients                                                                                                                         \\ \hline
$N$           & The total number of clients                                                                                                                \\ \hline
$q_i$         & \begin{tabular}[c]{@{}l@{}}The maximum number of client $i$ ’s CPU cycle per\\ second\end{tabular}                                       \\ \hline
$q_{ti}$      & The number of CPU cycles per second used to train                                                                                           \\ \hline
$q_{mi}$      & The number of CPU cycles per second used to mine                                                                                            \\ \hline

$p_{ti}$      & The unit price for training to client $i$                                                                                       \\
\hline
$p_{mi}$      & The unit price for mining to client $i$                                                                                        \\
\hline

$\pi$         & \begin{tabular}[c]{@{}l@{}}The number of training iterations for clients during \\ one round of BCFL to submit model update\end{tabular} \\ \hline
$D_i$         & The data size of client $i$                                                                                                  \\ \hline
$d_i$         & \begin{tabular}[c]{@{}l@{}}The number of CPU cycles used for training each\\ data sample\end{tabular}                                      \\ \hline
$\mu_i$       & \begin{tabular}[c]{@{}l@{}}The total CPU cycles required to finish the local \\ training for generating model updates\end{tabular}            \\ \hline
$T_i$         & The time spent on training for client $i$                                                                                                  \\ \hline
$\psi$        & The total CPU cycles used  to  mine  for  each  client                                                                                     \\ \hline
$T_{mi}$      & The time spent on  mining for client $i$                                                                                                   \\ \hline
$U_i$         & The  utility  of  client $i$ in  one  round of BCFL                                                                                        \\ \hline
$U_{mo}$      & The utility of the MO in one round of BCFL                                                                                                 \\ \hline

$q_{ti}^*$      & The optimal CPU cycles per second for training                                                                                           \\ \hline
$q_{mi}^*$      & The optimal CPU cycles per second for mining                                                                                            \\ \hline

$p_{ti}^*$      & The optimal unit price for training to client $i$                                                                                       \\
\hline
$p_{mi}^*$      & The optimal unit price for mining to client $i$                                                                                        \\
\hline
\end{tabular}
\label{table:not}
\end{table}

\subsection{System Overview}
Inspired by \cite{wang2021blockchain}, we consider a fully coupled BCFL system, which runs FL on a consortium blockchain network. In such a decentralized BCFL system, the participants in FL work as the blockchain nodes as well. Specifically, there are multiple local devices, termed as clients, working collaboratively to train a machine learning model, i.e., the global model. The set of clients can be denoted as $\mathcal{N}=\left\{ 1,\cdots,i,\cdots,N\right\}$ with $N$ representing the total number of clients in the BCFL system. 
For simplicity, we refer to the work done by the blockchain for FL as \textit{mining} in a uniform way, which does not imply that clients perform mining jobs consuming excessive amount of computing power like Proof of Work (PoW) \cite{nakamoto2008bitcoin}. In our system, we consider that lightweight consensus algorithms, such as Practical Byzantine Fault Tolerance (PBFT) \cite{castro1999practical} and Delegated Proof of Stake (DPoS) \cite{mingxiao2017review}, are utilized in the consortium blockchain system.

In our considered BCFL system, each client should be responsible for both training and mining. Since the workflow of the fully coupled BCFL is that mining starts only after the training is completed, we assume that training and mining are not parallel in this paper.
Once the training is finished, all the clients will upload their local model updates to the blockchain network so as to be recorded on the blockchain.
Here we define one round of BCFL as finishing both the training and mining processes.
Since mobile devices usually have their own tasks to finish rather than only contributing to BCFL, we assume that they will not use all of their computation resources. In other words, CPU cycles per second for training and mining can be adjusted strategically in each round of BCFL. 

\begin{figure}
\centering
\includegraphics[width=0.4\textwidth]{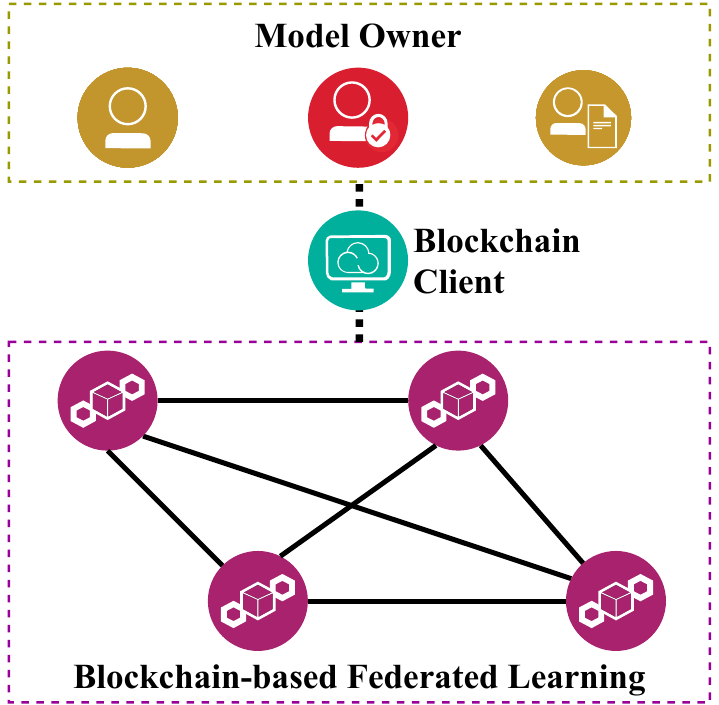}
\caption{An illustration of the BCFL system.}
\label{fig:bcfl}
\end{figure}

The MO is the requester of the FL task, aiming at to receive a well-trained final global model from the BCFL system. After the FL task is published on the blockchain, clients start to train their local models and then broadcast the obtained model updates to the blockchain network once the local training process is finished. By this means, the MO can only access the model updates from all clients rather than raw data of devices, thus preventing the leakage of private information for participants. An illustration of our system model is shown in Fig. \ref{fig:bcfl}. 

The detailed workflow to finish one BCFL task is as below:
\begin{itemize}
    \item The MO publishes an FL task, with the rewards for training and mining.
    \item Clients determine the computational resources, i.e., CPU cycles per second, used to train the model and mine for the blockchain based on the rewards provided by the MO.
    \item Each client trains the local model, and then broadcasts the model updates to the blockchain network. Then, clients start to mine the block.
    \item Once the block is generated, the model updates are stored on the blockchain, and the rewards will be delivered to each client.
    \item 
    Clients calculate the global model with the verified model updates on chain. As long as the expected performance of the global modal is not reached, clients will start the next round of training based on the aggregated global model.
    
\end{itemize}

\subsection{Utility Models}
Since the computational resource of client $i$ is limited  and we assume that client $i$ has other tasks to finish rather only working for the BCFL,  it is essential to design a decision  mechanism for client $i$ to allocate CPU cycles for training and mining, respectively. What's more, an incentive mechanism is necessary because clients will be reluctant to contribute to the BCFL task without receiving enough compensation for their efforts. It is difficult to design such a mechanism because the MO has limited rewards budget, and it is necessary to make sure that the time consumption of training and mining can be shortened and a well performed global ML model can be obtained at the same time. Thus, in the following part, we build the utility models of both the MO and client from the perspectives of resource allocation and incentive mechanism.

\subsubsection{Client's Utility}\label{sec:client's utility}

We assume that the maximum number of client $i$'s CPU cycles per second is $q_i$, and the number of CPU cycles per second used to train and mine are $q_{ti}$ and $q_{mi}$, respectively. Then we have $q_{ti}, q_{mi} \leq q_i$. 
Let $\pi$ be the number of training iterations for clients during one round of BCFL to submit model update, which is usually fixed for all clients. 
Let $D_i$ be the number of the data size of client $i$, and $d_i$ be the number of CPU cycles used for training each data sample. Therefore, we can define the total CPU cycles required to finish the local training to generate model updates as $\mu_i = \pi d_i D_i$.


Since any client $i$ can decide its CPU cycles used to train the local model, the time used to finish the local training varies for each client. We can calculate the time spent on training for client $i$ via $T_{ti} = \frac{\mu_i}{q_{ti}}$.
Besides, we denote the total CPU cycles used to mine for each client as $\psi$, which is the same for all clients since mining a new block in blockchain system usually consumes fixed computational resources. Thus, the time spent on mining can be calculated as $T_{mi}= \frac{\psi}{q_{mi}}$.
So we can have the total time cost of client $i$ to finish a round of BCFL task as $T_i = T_{ti}+T_{mi}$.
Since it is not possible to let $T_{ti}$ and $T_{mi}$ be limitless according to the convergence time requirement, we denote the upper bound of time consumption in one round of BCFL by $T$. Thus, we have $T_i \leq T$. 


In order to encourage clients to join BCFL, the MO provides some rewards to clients, where the prices per second for training and mining are denoted as $p_{ti}$ and $p_{mi}$, respectively. Clients can allocate unit CPU cycles for training and mining based on the unit prices given by the MO.
Then the rewards of client $i$ for training and mining to generate one round of local model updates are $R_{ti} = T_{ti}  p_{ti}$ and $ R_{mi} = T_{mi}p_{mi}$, respectively.
Thus, the total rewards for client $i$ in one round of BCFL is $R_i = R_{ti} + R_{mi}$.

Next, we can calculate the energy costs for training and mining as $C_{ti} = \rho_i \mu_i q_{ti}^2$ and $C_{mi} = \rho_i \psi q_{mi}^2$ base on a widely used energy consumption model\cite{burd1996processor}, where $\rho_i$ is the parameter correlated to the chip architecture.

In this way, the total cost\footnote{As for the communication cost, since the submissions of clients are the same, we can consider it as a constant value, which cannot be optimized anymore and thus is omitted here.} can be calculated as $C_i=C_{ti}+C_{mi}$. 

Finally, we can obtain the utility of client $i$ in one round of BCFL as
\begin{align}
    U_i & = R_i-C_i \notag \\
    &=\frac{\mu_i}{q_{ti}}p_{ti}+\frac{\psi}{q_{mi}} p_{mi}-\rho_i \mu_i q_{ti}^2-\rho_i \psi q_{mi}^2 .
    \label{client_u}
\end{align}

\subsubsection{MO's Utility}\label{mo_utility}
The main concerns related to the utility of the MO are the performance of the global model, the time consumption and the rewards paid to all participants in each round of BCFL, where the first one is a sort of revenue and the last two are related to the cost for the MO.

Generally, the performance of an ML model will be affected by the number of CPU cycles spent for training. Thus, we define the performance of the global model after one round of local training and mining as $G$ which can be calculated by $G=f(\sum_{i=1}^{N} \mu_i)$. Here $f(\cdot)$ is a monotonically increasing function, indicating that more CPU cycles used for the local training by all clients, the better performance of the global model after aggregation will be achieved.
As for the MO, its utility depends on the performance of the BCFL system ($G$), total time cost ($\sum_{i=1}^{N}(T_i)$), and total rewards distributed to clients ($\sum_{i=1}^{N}(R_i)$).
Thus, the utility of the MO in one round of BCFL can be expressed as 
\begin{align}
    U_{mo} & =f \left(\sum_{i=1}^{N} \mu_i\right)-\xi \sum_{i=1}^{N}(T_i+R_i) \notag \\
    & =f \left(\sum_{i=1}^{N} \mu_i\right)-\xi \sum_{i=1}^{N} \left(\frac{\mu_i}{q_{ti}}+\frac{\psi}{q_{mi}}+\frac{\mu_i}{q_{ti}} p_{ti}+\frac{\psi}{q_{mi}} p_{mi}\right), 
    \label{eq:u_mo}
\end{align}
where $\xi>0$ is a scalar parameter to balance the revenue and cost.


\subsection{Problem Formulation using Two-stage Stackelberg Game}
According to the above analysis of our system model, client $i$ provides its computational power to finish BCFL tasks based on the rewards given by the MO. In other words, the unit prices $p_{ti}$ and $p_{mi}$ determine the unit computational power $q_{ti}$ and $q_{mi}$. We can formulate the interactions between clients and the MO as a two-stage Stackelberg game, which is widely used for the complete information dynamic game \cite{zhang2009stackelberg}. In this game, the MO determines the unit prices of the CPU-cycle frequency used for training and mining, and then client $i$ decides its CPU cycles per second based on the received prices, which means that the decision of client $i$ is impacted by the decision of the MO. In this case, we can define the process of the two-stage Stackelberg game as below:
\begin{itemize}
    \item Stage \uppercase\expandafter{\romannumeral1}: The MO sets the unit prices per second for training and learning for each client, i.e., $p_{ti}$ and $p_{mi}$, via maximizing its own utility, which is specifically based on its budget and the total number of CPU cycles consumed for training submitted by each client.  Taking into account the fairness of the distribution for setting price, we need to design a fair reward allocation scheme here.
    \item Stage \uppercase\expandafter{\romannumeral2}: After  receiving the unit prices from the MO, clients determine their corresponding computational power, i.e., $q_{ti}$ and $q_{mi}$, through optimizing their respective utilities. 
\end{itemize}



In practical situations, $q_{ti}$ and $q_{mi}$ are not independent of each other because of time and reward budget constraints; similarly, $p_{ti}$ and $p_{mi}$ influence each other as well. Therefore, we should take these constraints into account when modeling to make the decisions reasonable.

Recall $\mu_i$ in Section \ref{sec:client's utility}, we know that $\mu_i$ is a variable correlated to the data size of client $i$ and the performance of the corresponding device, which may not always known to the MO or the system. As for $\psi$, it can be predefined by the system since generating a new block usually consumes a constant amount of resources. Therefore, we can classify the two-stage Stackelberg game into information complete and incomplete scenarios based on whether $\mu_i$ is known to the MO. The models derived for these two scenarios are different, and hence the strategies of the MO and clients are different accordingly, which will be explored in Sections \ref{sec:game1} and \ref{game2}.

\section{Resource Allocation with Complete Information}\label{sec:game1}

In this section, we will elaborate the expressions of the proposed Stackelberg game model and the corresponding solutions for the clients and the MO in the scenario of complete information, which means that the MO makes its decisions when $\mu_i$ of each client is known as a prior. First, we propose a fair reward allocation scheme for clients, and then we transfer the two-stage Stackelberg game into two separate optimization problems that are resolved sequentially. The methodology we adopt to solve the two problems is backward induction, which requires analyzing the optimal strategies of Stage \uppercase\expandafter{\romannumeral2} first and then the strategies of Stage \uppercase\expandafter{\romannumeral1}.

\subsection{Fair Reward Allocation}

Before we formulate the game model, we should clarify the fair reward allocation scheme first.
In our system, we consider that each client has an equal chance to participate in both the training and mining processes with fair rewards. 
And since the allocation of rewards to each client in training and mining has a significant impact on the system fairness and further the participation willingness, we need to design a fair reward allocation scheme.
Although we have already defined the payoff of each client during the training and mining processes in the above section, it is necessary to investigate their upper bounds based on the MO's rewards budget. And the rewards distribution should not only be associated with the computing power of the device, but also take into account the performance of its work. On the one hand, the reward budget of the MO and the rewards that each client can get are limited; on the other hand, if the resources are allocated only based on the computing power devoted, it could lead to the situation where devices with sufficient computing power take most of the rewards, while devices with less power cannot get enough rewards, making the system unstable and unsustainable.

\subsubsection{Upper Bound of Rewards for Mining}

For simplicity, we set a fixed total reward budget $\eta$ in each round of BCFL. Since the computational power consumed by generating a new block is constant, with $\eta_m$ denoting the upper bound of the reward for mining that all clients can receive, we have:
\begin{align}
    \frac{\psi}{q_{mi}}p_{mi} \leq \overline{R}_{mi}= \frac{\eta_m}{N},
    \label{r_mi}
\end{align}
where $\overline{R}_{mi}$ is the upper bound of the reward for mining that each client can get.
\subsubsection{Upper Bound of Rewards for Training} 
Since the devices in our BCFL system are assumed to be homogeneous, and they may have different computing capabilities,
we cannot simply distribute the rewards evenly to each client. To guarantee the fairness of reward distribution, we allocate rewards based on the contribution of each client in the training process. Considering that Shapely Value (SV) \cite{qu2020privacy} is a methodology which can distribute the rewards to participants according to their respective contributions, here
we apply it to facilitate reward distribution. 
The SV of client $i$ is defined as
\begin{align}
    SV_i(\mathcal{N},v) = \sum_{i\notin S, S\subseteq \mathcal{N}}{\frac{s!(N-s-1)!}{N!}}(v(S\cup i)-v(S)),
    \label{eq:sv}
\end{align}
where 
$S \subseteq \mathcal{N}$ is a subset of clients and $s=| S |$ is the number of devices in the set $S$; $v(S)$ is a function describing the performance of the training result with the client set $S$. Then, we give the expression of function $v(S)$. Recall $G$ in Section \ref{mo_utility}, we can assume that $v(S)$ is a function correlated to $G$ and it can be defined as
\begin{align}
    v(S)=W-\left \| \frac{\sum_{i=1}^{s}{G}}{s}-g\right \|_2,
    \label{eq:vs}
\end{align}
where $W=max_{S\subseteq \mathcal{N}}\left \| \frac{\sum_{i=1}^{s}{G}}{s}-g\right \|_2$ and $\left \|\cdot \right \|_2$ is the Euclidean norm; $g$ is the targeted performance value. Then, we can calculate the upper bound of the reward distributed to each device for training as:
\begin{align}
\overline{R}_{ti}=\frac{SV_i(\mathcal{N},v)}{v(\mathcal{N})}(\eta-\eta_m).
\end{align}

For each client, its rewards should not exceed the upper bound, so we can have the following constraint:
\begin{align}
    \frac{\mu_i}{q_{ti}}p_{ti}\leq \overline{R}_{ti}.
    \label{eq:r_ti}
\end{align}


\subsection{Stage \uppercase\expandafter{\romannumeral2}: Clients Set CPU Cycles Per Second based on Unit Rewards}\label{se:stage2}
Since each client $i$ has a limited amount of computational resource and should follow the working rules of BCFL, the goal of client $i$ is to maximize its utility as follows:
\begin{align}
\textbf{Problem 1:} \quad \max: \quad & U_i, \notag \\
s.t.\quad &\frac{\mu_i}{q_{ti}} p_{ti}-\rho_i \mu_i q_{ti}^2 \geq 0, \label{eq:prob1_constraint1}\\
&  \frac{\psi}{q_{mi}}  p_{mi}-\rho_i \psi q_{mi}^2 \geq 0, \label{eq:prob1_constraint2}\\
& \frac{\mu_i}{q_{ti}}+\frac{\psi}{q_{mi}}\leq T, \forall i\in \mathcal{N}, 
\label{eq:prob1_constraint3}
\end{align}
 where the first two constraints \eqref{eq:prob1_constraint1} and \eqref{eq:prob1_constraint2} mean that client $i$ wishes to gain non-negative payoffs in both training and mining; and the last constraint \eqref{eq:prob1_constraint3} indicates that client $i$ should finish the working process, including training and mining, within the time period $T$. 
 
 

It is clear that Problem 1 is a nonlinear optimization problem with inequality constraints, so we adopt the method of Karush-Kuhn-Tucker (KKT) conditions to solve it. First, we need to demonstrate that Problem 1 can be resolved. We can calculate $\frac{\partial U_i}{\partial q_{ti}}=-\frac{\mu_i p_{ti}}{q_{ti}}-2 \rho q_{ti}<0$ and $\frac{\partial U_i}{\partial q_{mi}}=-\frac{\mu_i p_{mi}}{q_{mi}}-2 \rho q_{mi}<0$, so it can be proved that $U_i$ is concave and it has the maximum value. By solving Problem 1, we get the following theorem:
\begin{theorem}\label{eq:df2}
The optimal strategies of client $i$ in the scenario of complete information are given by
\begin{align}
q_{ti}^* &  = \left({\frac{p_{ti}}{\rho_i}}\right)^\frac{1}{3},
\label{eq:q_ti_1}\\
q_{mi}^* & =\frac{\psi}{T- \mu_i \left(\frac{\rho_i}{p_{ti}}\right)^\frac{1}{3}}.
\label{eq:q_m}
\end{align}
\end{theorem}
The detailed proof of Theorem \ref{eq:df2} is in Appendix A. From the above theorem, we can see that the number of optimal CPU cycles per second client $i$ putting into training grows as the unit price for training given by the MO increases. The optimal CPU cycles per second devoted to mining is constraint by $\psi$, indicating that if the mining work requires more CPU cycles, client $i$ should mine with a larger $q_{mi}^*$.

\subsection{Stage \uppercase\expandafter{\romannumeral1}: MO Sets Unit Prices for Clients}

The MO expects to get a global model with good performance consuming time and cost for rewards as less as possible, so its goal is to maximize the utility function $U_{mo}$, and the optimization problem can be formulated as follows:
\begin{align}
\textbf{Problem 2:} \quad \max: \quad & U_{mo}, \notag\\
s.t.\quad &\frac{\mu_i}{q_{ti}}p_{ti}\leq \overline{R}_{ti}, \label{con_r_ti}\\
&\frac{\psi}{q_{mi}}p_{mi}\leq \overline{R}_{mi}, \forall i\in \mathcal{N}, \label{con_r_mi}
\end{align}
where 
(\ref{con_r_ti}) and (\ref{con_r_mi}) are the constraints of individual rewards from training and mining to meet the MO's budget. 

It is clear that Problem 2 is also a nonlinear optimization problem, and $U_{mo}$ is also concave, so we can list all the KKT conditions to find its maximum value. Via solving Problem 2, we can have:

\begin{theorem}
\label{def 2}
The optimal strategies of the MO in the scenario of complete information are:
\begin{align}
    p_{ti}^*& =\left(\frac{1}{\rho_i}\right)^\frac{1}{2}
    \left({\frac{\overline{R}_{ti}}{\mu_i}}\right)^\frac{3}{2},
    \label{eq:optimal_p_t} \\
    p_{mi}^*& =\frac{\overline{R}_{mi}}{T-(\rho_i \mu_i)^\frac{3}{2} \left(\frac{1}{\overline{R}_{ti}}\right)^\frac{1}{2}}.
    \label{eq:optimal_p_m}
\end{align}



\end{theorem}

The proof of Theorem \ref{def 2} is shown in Appendix B. In the optimal solutions above, $p_{mi}^*$ and $p_{ti}^*$ are highly correlated. This is because there are time and budget constraints so that $p_{ti}^*$ and $p_{mi}^*$ are not independent variables from each other. In other words, the MO needs to balance $p_{ti}^*$ and $p_{mi}^*$ to satisfy the constraints when making decisions. Furthermore, we can find that $\mu_i$ and $\psi$ influence the optimal decisions as well.

We summarize the resource allocation mechanism with complete information in Algorithm \ref{al:complete}. The MO calculates the unit prices given to the client for training and mining first, and then calculates its utility based on the previous unit prices (Lines 1-2). If $U_{mo}$ is the optimal utility for the MO, then the optimal decisions of MO can be obtained (Lines 3-5). Next, the MO sends the unit prices to clients, and each client calculates the the numbers of CPU cycles per second used for training and mining; if the utility for client $i$ is optimal, client $i$ can make its optimal decisions and start to train and mine (Lines 6-12).

\begin{algorithm}
\caption{Resource Allocation Mechanism with Complete Information} 
\label{al:complete}
\begin{algorithmic}[1]
\REQUIRE $T$, $\mu_i$, $\psi$, $\rho_i$, $\eta$, $\overline{R}_{mi}$
\ENSURE $q_{ti}^*$, $q_{mi}^*$, $p_{ti}^*$, $p_{mi}^*$

\STATE The MO calculates $\hat{p}_{ti}$ and $\hat{p}_{mi}$ via (\ref{eq:optimal_p_t}) and (\ref{eq:optimal_p_m})
\STATE The MO calculates $U_{mo}$ based on $\hat{p}_{ti}$ and $\hat{p}_{mi}$ via (\ref{eq:u_mo})
\IF{$U_{mo}(\hat{p}_{ti},\hat{p}_{mi})\geq U_{mo}(p_{ti},p_{mi})$}
\STATE $p_{ti}^*\leftarrow \hat{p}_{ti}$, $p_{mi}^* \leftarrow \hat{p}_{mi}$
\ENDIF
\STATE The MO sends $p_{ti}^*$ and $p_{mi}^*$ to the client $i$
\FOR{$i \in \mathcal N$}
\STATE Calculate  $\hat{q}_{ti}$ and $\hat{q}_{mi}$ via (\ref{eq:q_{ti}_1}) and (\ref{eq:q_m})
\IF{$U_i(\hat{q}_{ti}, \hat{q}_{mi}) \geq U_i(q_{ti}, q_{mi})$}
\STATE $q_{ti}^*\leftarrow \hat{q}_{ti}$, $q_{mi}^* \leftarrow \hat{q}_{mi}$
\STATE Client $i$ uses $q_{ti}^*$ to train and $q_{mi}^*$ to mine
\ENDIF
\ENDFOR
\RETURN $q_{ti}^*$, $q_{mi}^*$, $p_{ti}^*$, $p_{mi}^*$
\end{algorithmic}
\end{algorithm}

In general, the case of complete information is an ideal situation, and we find that it mainly influences the optimal decisions of the MO. Therefore, we can study the optimal decisions in the case of incomplete information by adjusting the decision mechanism of the MO.

\section{Resource Allocation with Incomplete Information}\label{game2}
In this section, we will discuss the game model in the case of incomplete information where the MO has no knowledge of the true value of $\mu_i$ for each client. Thus, the MO needs to set the unit price in such a way that each client has a non-negative payoff, while  ensures that the clients report the  value of $\mu_i$ honestly. Before designing the resource allocation mechanism, we first give two definitions below.

\begin{definition}(Individual Rationality). The incentive mechanism for resource allocation is individually rational if the utility of client $i$ given the rewards provided by the MO is non-negative, i.e., 
\begin{align}
    U_i(q_{ti},q_{mi},p_{ti},p_{mi}, \mu_i)\geq 0, \forall i.
    \label{df_ir}
\end{align}
\end{definition}
\begin{definition}(Incentive Compatibility). The incentive mechanism for resource allocation is incentive compatible if each client can get the optimal utility by reporting its $\mu_i$ truthfully, i.e.,
\begin{align}
    U_i(q_{ti},q_{mi},p_{ti},p_{mi},\mu_i)\geq U_i(q_{ti},q_{mi},p_{ti},p_{mi},\hat{\mu}_i),\forall i,
    \label{df_ic}
\end{align}
where $\hat{\mu}_i$ represents any value of $\mu_i$.
\end{definition}

Based on the previous analysis, we know that clients' decisions are made based on their non-negative utility. Since clients should ensure that the rewards they receive are not less than the total costs they spend, in such a situation, they can participate in the BCFL task. So in the situation of incomplete information, the MO needs to guarantee that its decisions should satisfy (\ref{df_ir}) to encourage clients to join the work. Besides, $\mu_i$ of client $i$ is not known by the MO, and the decisions of the MO are required to be based on the correct value of $\mu_i$ reported by clients, so the MO needs to satisfy (\ref{df_ic}) when making the decisions.

Since the client sets the CPU cycles per second after the unit prices are given by the MO, the decisions of the client in the case of incomplete information are the same as those made under the complete information case as discussed in Section \ref{se:stage2}. Therefore, we will only focus on the derivation of the optimal strategies of the MO in this section. 

With incomplete information, the MO has to ensure that the allocation of rewards to all clients is fair, the clients' utilities are non-negative, and clients report $\mu_i$ truthfully. Thus, the decision-making problem of the MO with incomplete information can be transformed into the following optimization problem:
\begin{align*}
    \textbf{Problem 3:} \quad \max: \quad & U_{mo}\\
s.t.\quad &  (\ref{con_r_ti}), (\ref{con_r_mi}), (\ref{df_ir}), (\ref{df_ic}),\\
& \forall i\in \mathcal{N},
\end{align*}
where (\ref{df_ir}) and (\ref{df_ic}) are the constraints of individual rationality and incentive compatibility for the mechanism; (\ref{con_r_ti}) and  (\ref{con_r_mi}) are the constraints of individual rewards for meeting the MO's budget. 

To solve Problem 3, we can first write it in Lagrangian form according to its optimization objective and constraints, and then analyze its KKT conditions. The optimal solutions can be solved as below: 
\begin{theorem}
The optimal strategies of the MO in the scenario of incomplete information are
\begin{align}
    p_{ti}^* & =\left(\frac{1}{\rho_i}\right)^\frac{1}{2}
    \left({\frac{\overline{R}_{ti}}{\mu_i}}\right)^\frac{3}{2},
    \label{eq:optimal_p_t_2}\\
     p_{mi}^* & =\frac{\rho_i \,\psi^3 }{\left(T-\mu_i \left(\frac{\rho_i}{p_{ti}^*}\right)^\frac{1}{3}\right)^3}
     \label{eq:optimal_p_m_2}.
\end{align}
\label{th_3}
\end{theorem}
The proof of Theorem \ref{th_3} is presented in Appendix C. The optimal solution for $p_{ti}^*$ in the incomplete information case is the same as the optimal solution in the complete information case, while $p_{mi}^*$ is different. Since the decision of the MO in the case of incomplete information is not only influenced by the budget of the reward, but also required to satisfy the two conditions  (\ref{df_ir}) and (\ref{df_ic}) in the above definitions. In other words, the decisions in this case is more conservative so the MO would prefer to minimize its cost by reducing the payments to training and mining. We will illustrate the specific differences about the decisions in the two scenarios through experiments in Section \ref{sec:experiment}.

The resource allocation mechanism in the incomplete-information case is presented in Algorithm \ref{al:incomplete}, which is similar to Algorithm \ref{al:complete}, except the decision process of the MO. In the scenario, the MO should make sure that its utility is optimal and the utility for each client is non-negative (Lines 2-7). 
\begin{algorithm}
\caption{Resource Allocation Mechanism with Incomplete Information} 
\label{al:incomplete}
\begin{algorithmic}[1]
\REQUIRE $T$, $\mu_i$, $\psi$, $\rho_i$, $\eta$, $\overline{R}_{mi}$
\ENSURE $q_{ti}^*$, $q_{mi}^*$, $p_{ti}^*$, $p_{mi}^*$
\STATE The MO calculates $\hat{p}_{ti}$ and $\hat{p}_{mi}$ via ($\ref{eq:optimal_p_t_2}$) and ($\ref{eq:optimal_p_m_2}$)
\IF{$U_{mo}(\hat{p}_{ti},\hat{p}_{mi})\geq U_{mo}(p_{ti},p_{mi})$}
\STATE The MO calculates the expected utility $\hat{U_i}$ of client $i$ 
\IF{$\hat{U_i}\geq 0$}
\STATE $p_{ti}^*\leftarrow \hat{p}_{ti}$, $p_{mi}^* \leftarrow \hat{p}_{mi}$
\ENDIF
\ENDIF
\STATE The MO sends $p_{ti}^*$ and $p_{mi}^*$ to client $i$
\FOR{$i \in \mathcal N$}
\STATE Calculate  $\hat{q}_{ti}$ and $\hat{q}_{mi}$
\IF{$U_i(\hat{q}_{ti}, \hat{q}_{mi})\geq U_i(q_{ti}, q_{mi})$}
\STATE $q_{ti}^*\leftarrow \hat{q}_{ti}$, $q_{mi}^* \leftarrow \hat{q}_{mi}$
\STATE Client $i$ uses $q_{ti}^*$ to train and $q_{mi}^*$ to mine
\ENDIF
\ENDFOR
\RETURN $q_{ti}^*$, $q_{mi}^*$, $p_{ti}^*$, $p_{mi}^*$
\end{algorithmic}
\end{algorithm}

We can see that the time complexity of both Algorithm \ref{al:complete} and Algorithm \ref{al:incomplete} is $O(n)$, which means that the time consumption of solving these two optimization problems increases with the number of clients linearly. Therefore, our proposed algorithms can work efficiently in practice.

\section{Experimental Evaluation}\label{sec:experiment}

In this section, we will conduct numerical experiments to verify and support our designed mechanism. We first clarify the experimental settings and then illustrate the results. We implement the simulations using Matlab 2019b in macOS 11.0.1 running on Intel i7 processor with 32 GB RAM and 1 TB SSD. 

\subsection{Experimental Setting}

In our experiments, we mainly focus on the impacts of four variables (i.e., $\mu_i$, $\psi$, $p_{ti}$ and $p_{mi}$) on our designed models under complete and incomplete situations. The basic setting for these simulations are slight different, and we will clarify the different parts of the settings in each experiment. For simplicity of calculation and presentation, we use GHz as the unit of CPU cycles per second and minute as the unit of time. We first set $\eta=1500$ and $\overline{R}_{mi}=5$.
Since we adopt SV to calculate the total rewards distributed to individual client and SV is correlated to the value of $\mu_i$ (see (\ref{eq:sv}) and (\ref{eq:vs})), we let $G=\frac{\sum_{i=1}^{N}\mu_i} {N}$. By running the algorithm of SV we can get the value of $\overline{R}_{ti}$ for each client.
The settings for other parameters are  $\rho_i=0.01$, $\xi=0.1$, $ g=10$ and $T=15$. 
Note that we conducted extensive experiments with other experimental settings, while we found that different values of the parameters would not influence the trends of the results.
So we only present the results of the experiments based on the above settings.

\subsection{Experimental Results}
\begin{figure}[htpt]
\centering
\subfigure{\includegraphics[width=0.23\textwidth]{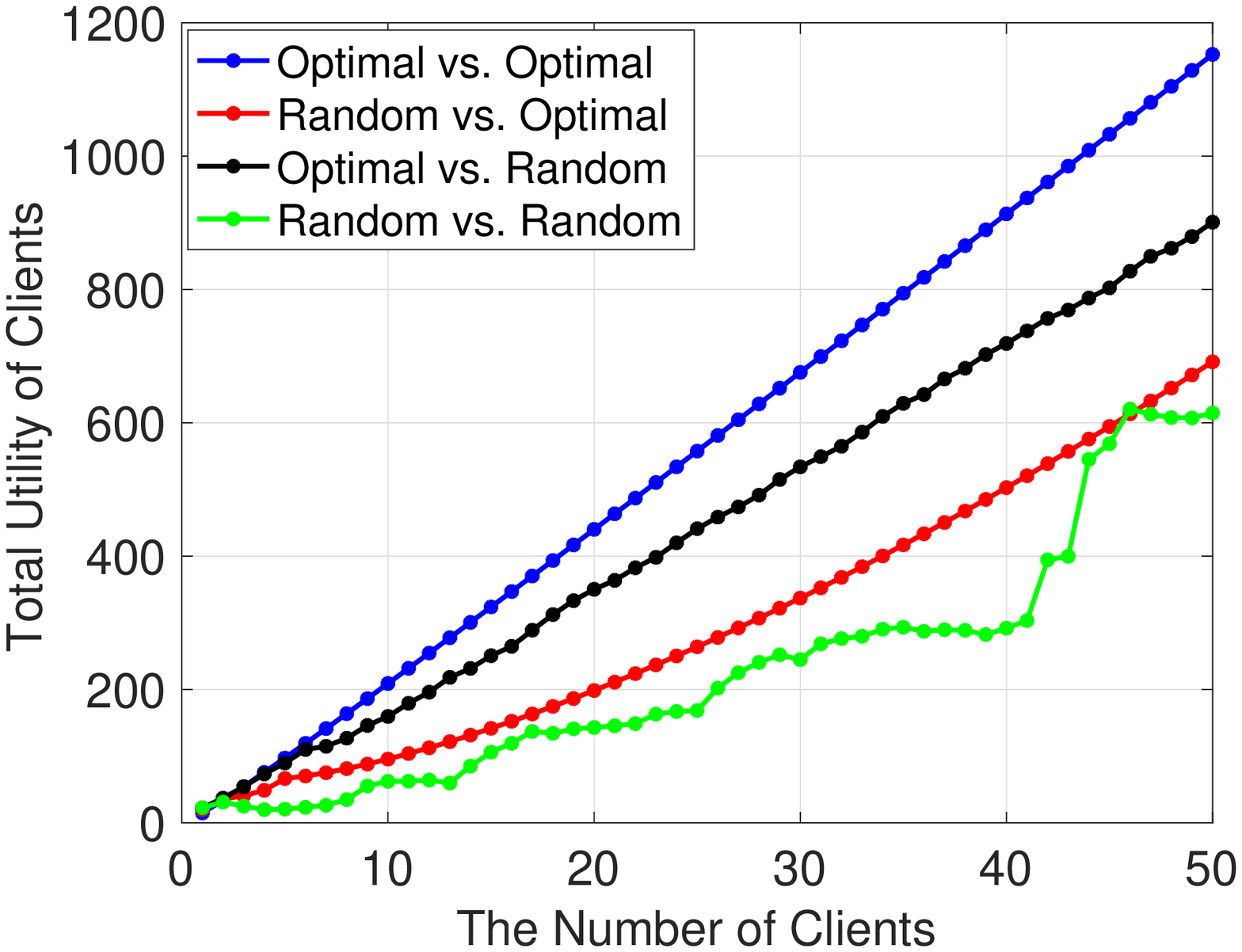}}
\subfigure{\label{fig:proof_solution_1} \includegraphics[width=0.23\textwidth]{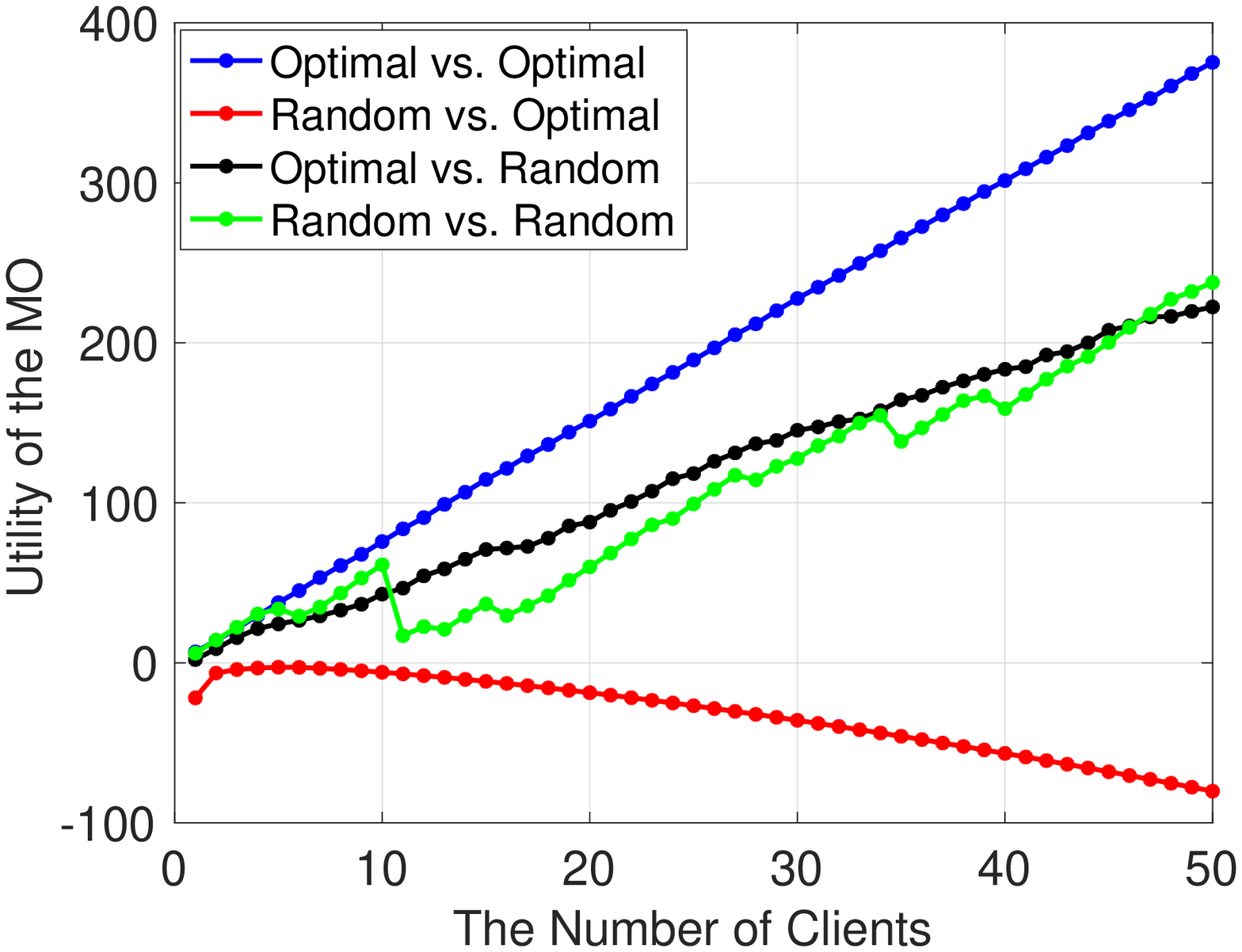}}
\caption{Utilities changing with strategy pairs.}
\label{fig:proof_solution_2}
\end{figure}

First, we prove the correctness of the optimal strategies derived from our models. We assume there are 50 clients in total and each client has the same data size, so we set $\mu_i=10$.  In our experiments, for clients and the MO, there are four strategy combinations, i.e., both sides choose the optimal strategies, one  chooses the random strategies while the other chooses optimal strategies, and both choose the random strategies. For example, we define the strategy combination \textsl{Random vs. Optimal} as the clients choose the random strategies and the MO chooses the optimal strategy.
We compare the utilities of clients and the MO with random strategies and optimal strategies, respectively. The results in Fig. \ref{fig:proof_solution_2} illustrate that clients and the MO can obtain the higher utilities than all other strategies when they both choose the optimal strategies, proving the validity of our proposed optimal strategies. 

\begin{figure}[htpt]
\centering
\subfigure{
\includegraphics[width=0.23\textwidth]{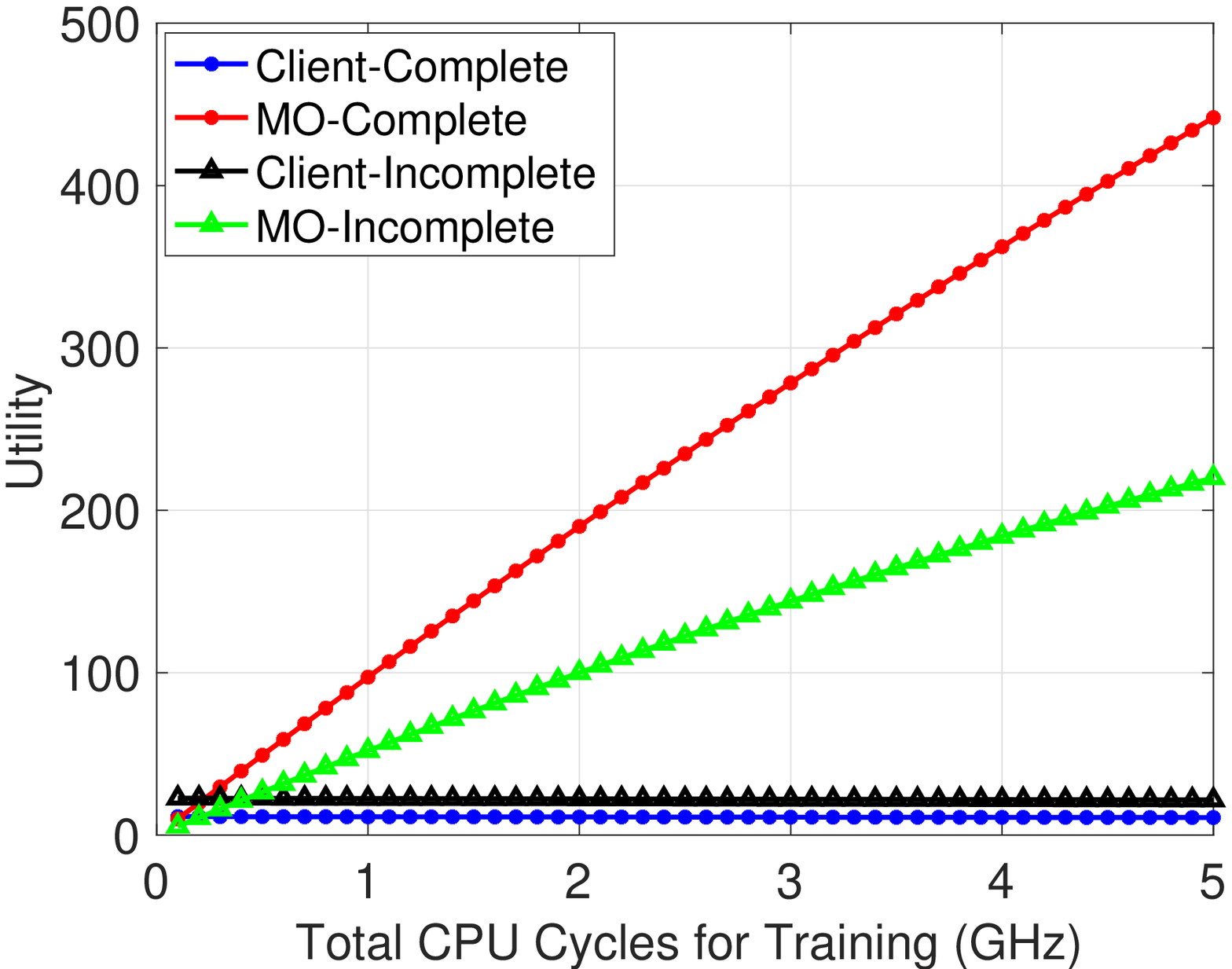}}
\subfigure{
\label{fig:clients}
\includegraphics[width=0.23\textwidth]{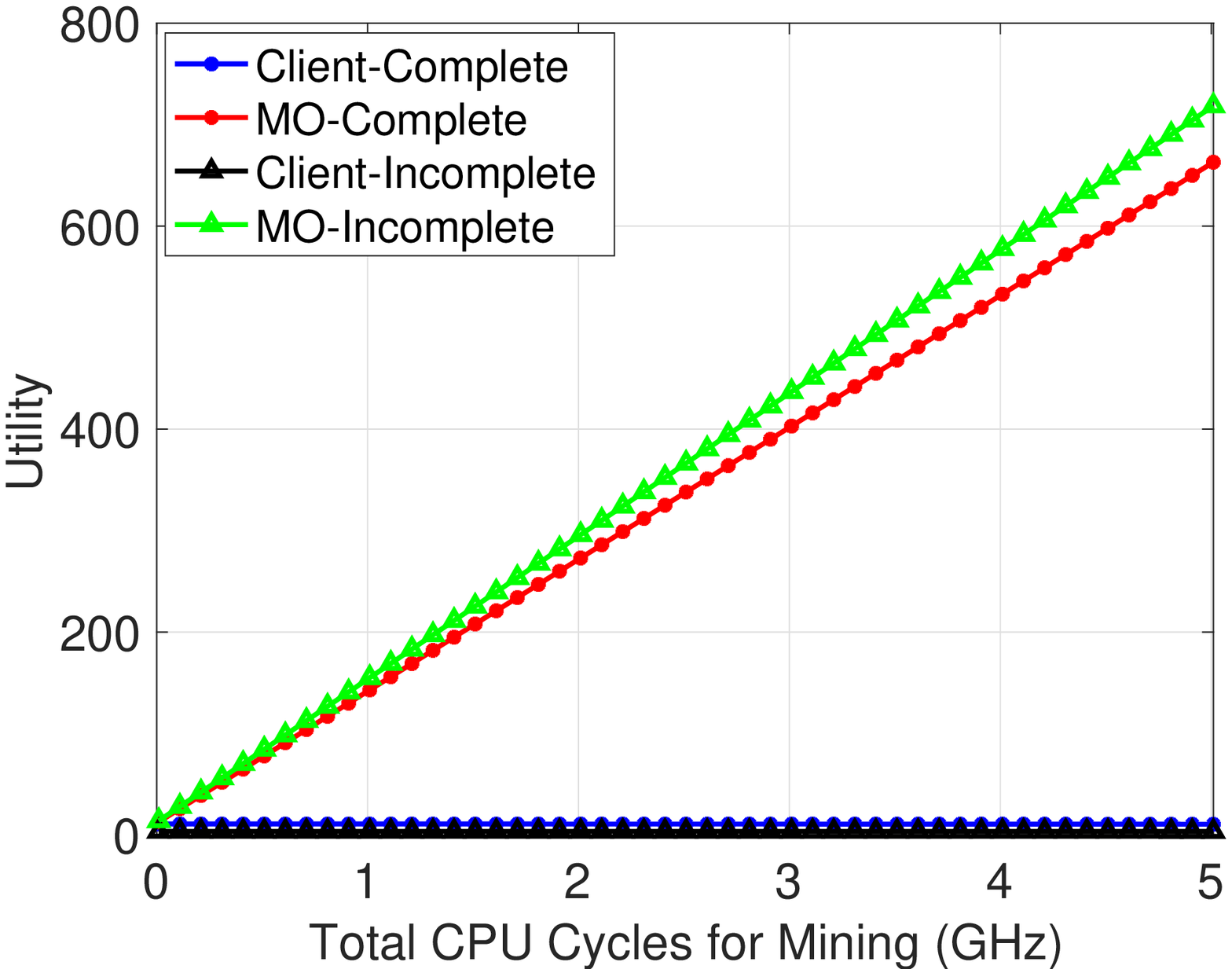}}
\caption{Utilities of the client and the MO changing with the total CPU cycles for training and mining.}
\label{fig:utility_u}
\end{figure}

Then, the experiments will be designed to study the impacts of $\mu_i$ and $\psi$ on the utility of clients and the MO under the situations of complete and incomplete information.
We set $\mu_i\in [0,5]$ and $\psi \in [0,5]$. The simulation results are shown in Fig. \ref{fig:utility_u}.
We can see that both $\mu_i$ and $\psi$ have a significant impact on the utility of the MO. That is because the higher CPU power will shorten the time in each round and improve the performance of the global model. However, for clients, devoting more CPU cycles does not result in more utility due to higher energy consumption. 

\begin{figure}[htpt]
\centering
\subfigure{
\includegraphics[width=0.23\textwidth]{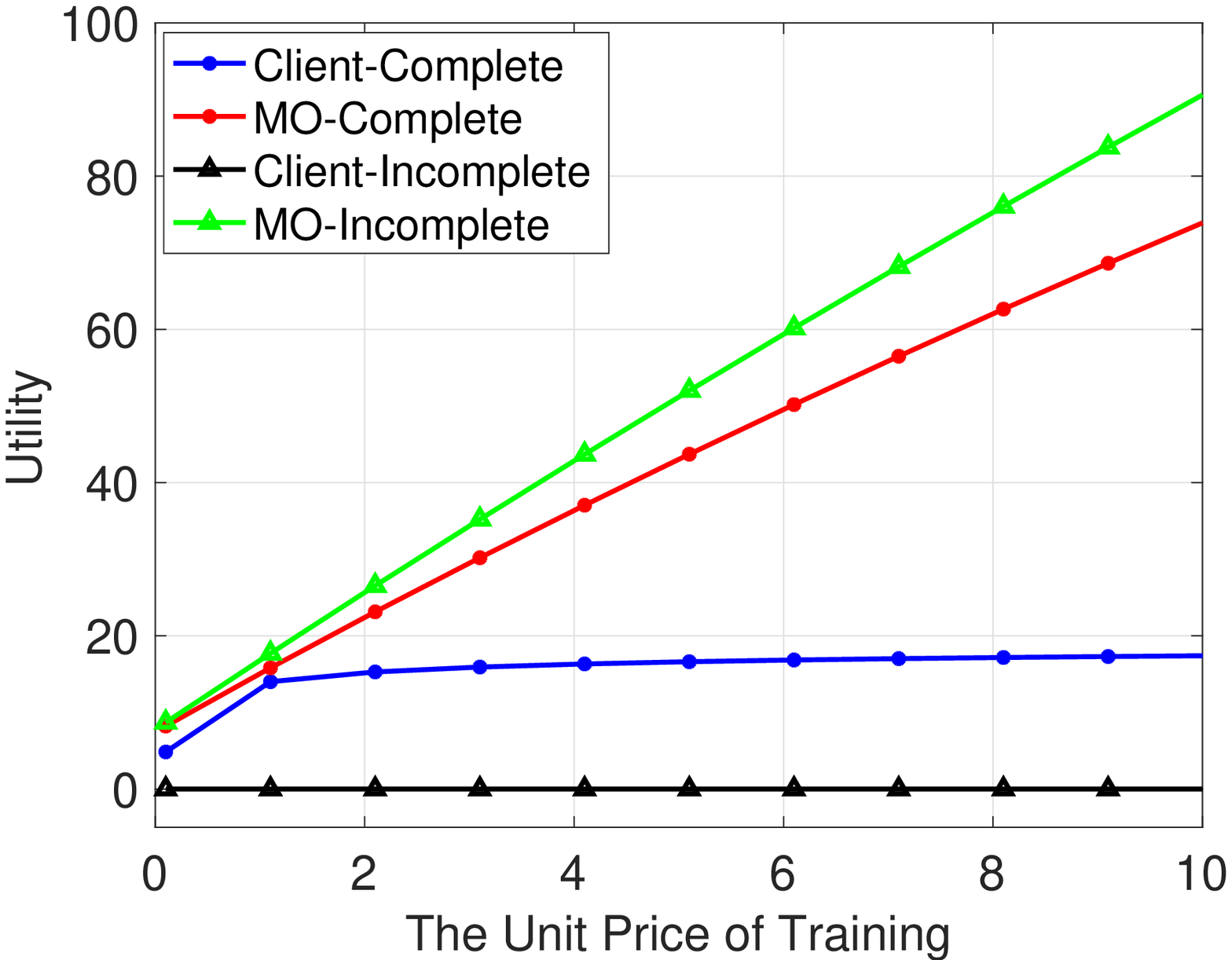}}
\subfigure{
\label{fig:clients}
\includegraphics[width=0.23\textwidth]{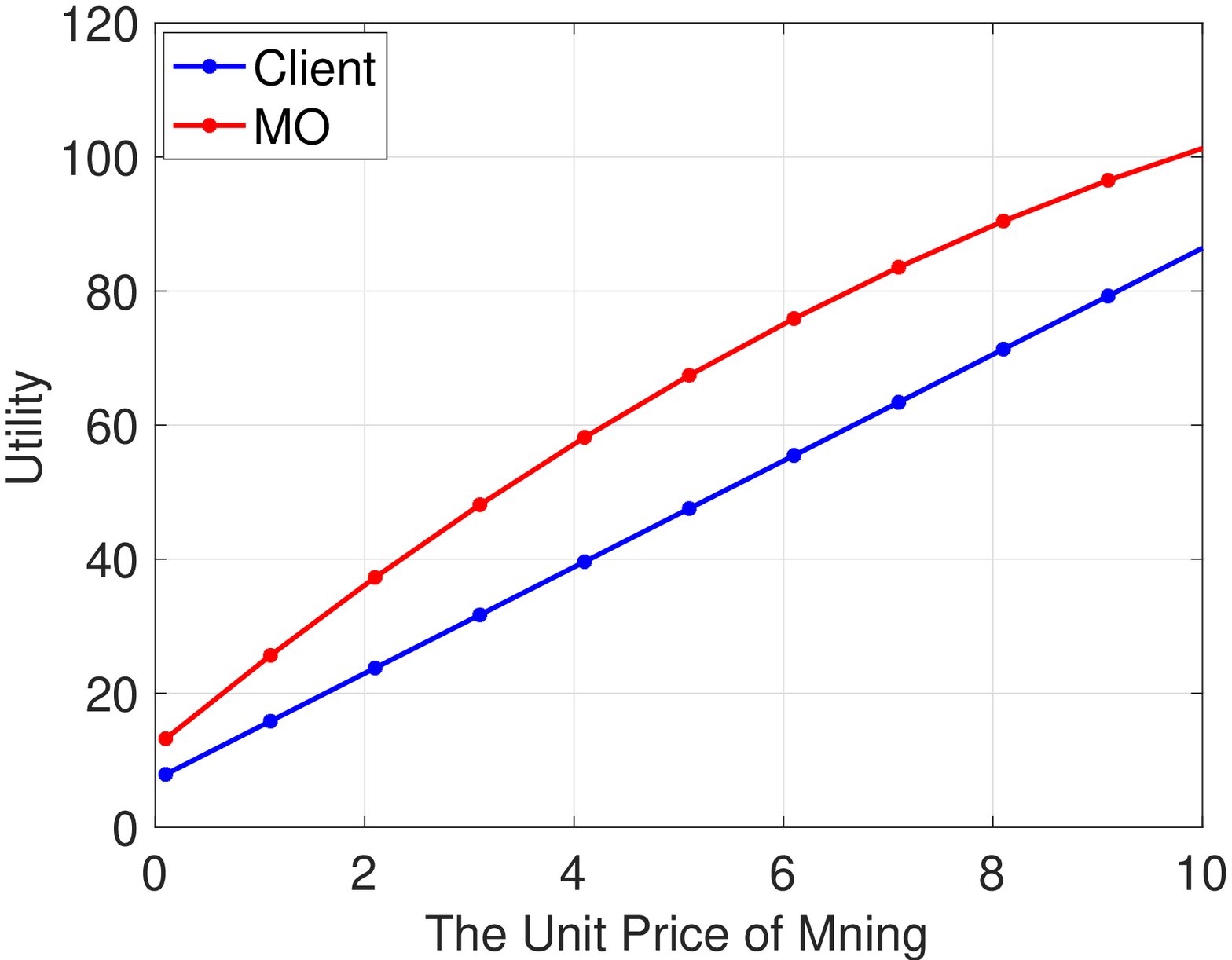}}
\caption{Utilities of the client and the MO changing with the unit prices of training and mining.}
\label{fig:utility_p}
\end{figure}

We then study the effect of $p_{ti}$ and $p_{mi}$ on the utility of the MO and clients. We set $p_{ti}\in [0, 10]$ and $p_{mi}\in [0, 10]$. The results are shown in Fig. \ref{fig:utility_p}. If the unit price of training increases, clients can be stimulated to provide more computing power, which reduces the time cost and improves the model performance, so the MO utility will be improved.  However, the revenue of clients does not grow significantly with the increase of the unit price of training, because the cost of energy consumption also rises.
$p_{mi}$ has the same effect on utility for both complete and incomplete information cases, and the results are shown on the right side of Fig. \ref{fig:utility_p}.
When the unit price of mining increases, the utility of both clients and the MO can be improved. This is because with the increase of $p_{mi}$, clients can receive more mining revenue by devoting more $q_{mi}$. At the same time, the MO can reduce the time cost and improve its utility by encourage clients to devote more CPU cycles per second for mining. 

\begin{figure}[htpt]
\centering
\subfigure{
\includegraphics[width=0.23\textwidth]{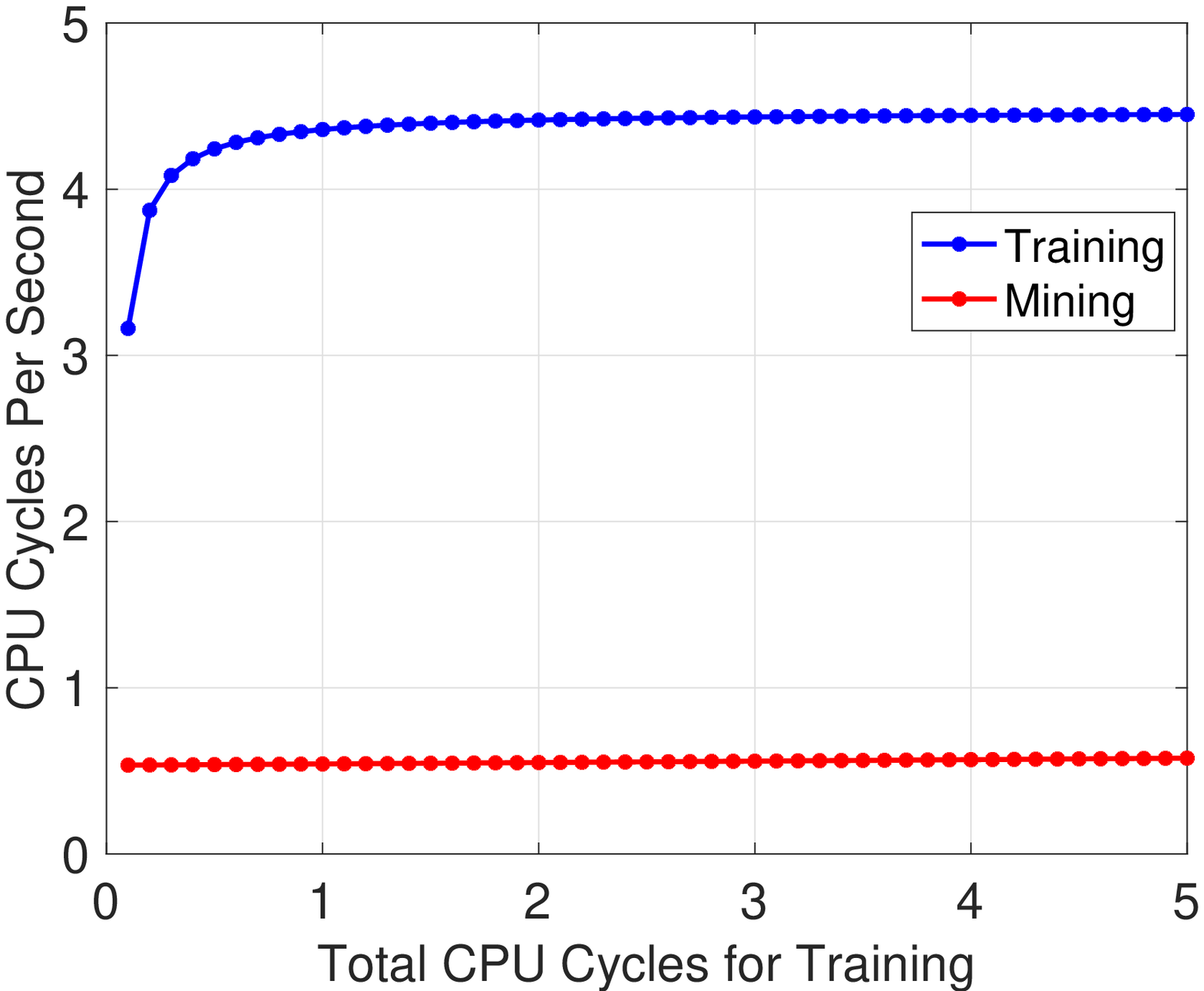}}
\subfigure{
\label{fig:clients}
\includegraphics[width=0.23\textwidth]{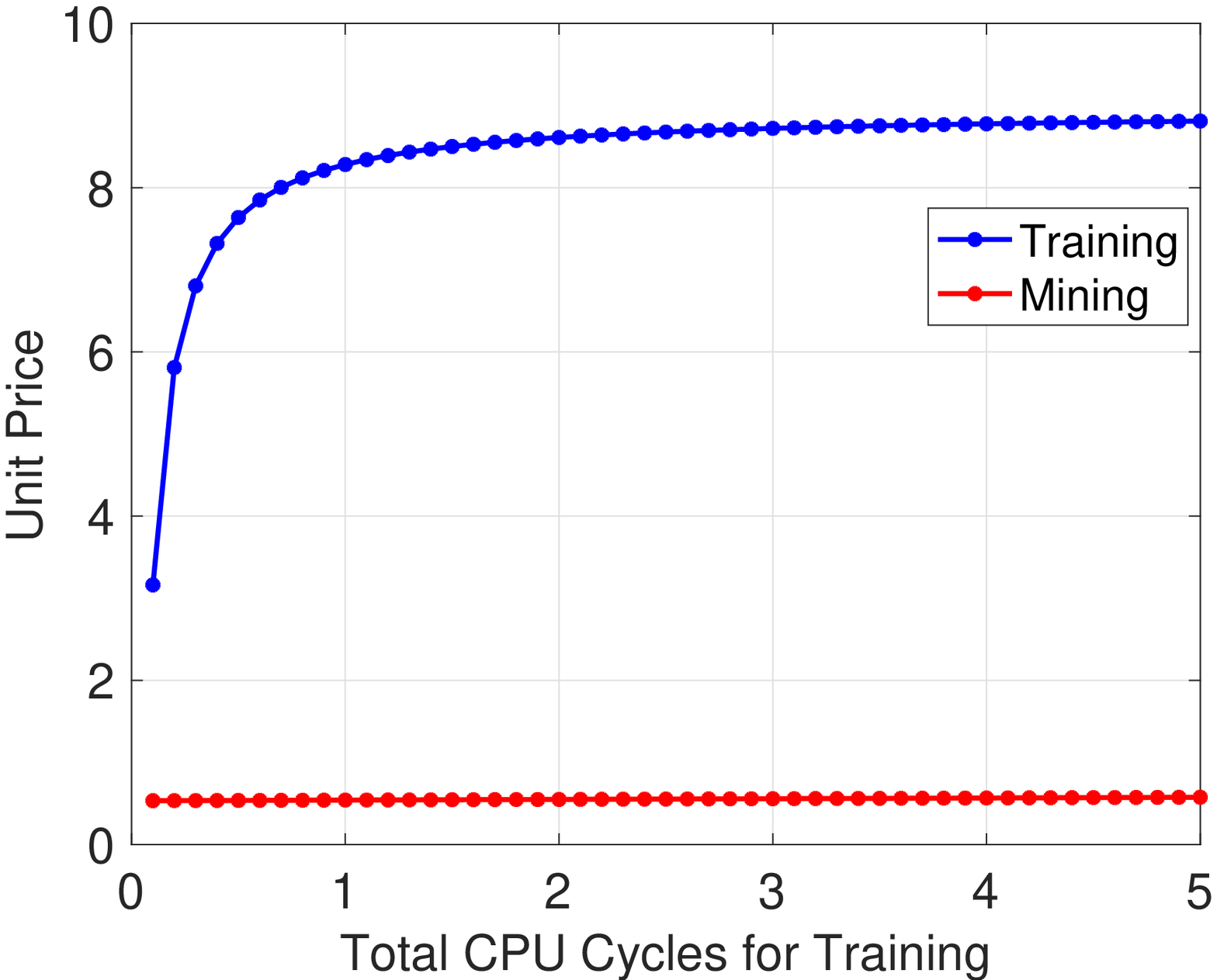}}
\caption{Impacts of $\mu_i$ on CPU cycles per second and unit prices for training and mining.}
\label{fig:u_q_i}
\end{figure}

Next, we conduct experiments to analyze the relationship between $\mu_i$ and the unit price for training and mining. We set $\mu_i\in[0,5]$, and the results are illustrated in Fig. \ref{fig:u_q_i}. We can see that both the unit price and the number of CPU cycles for training increase with $\mu_i$. This is because if $\mu_i$ increases, more rewards are needed to motivate clients to put more computational resources in training. In general, $\mu_i$ does not affect $p_{mi}$ and $q_{mi}$ a lot, as the benefits of mining are relatively constant and are more influenced by the resource allocation scheme.

\begin{figure}[htpt]
\centering
\includegraphics[width=0.33\textwidth]{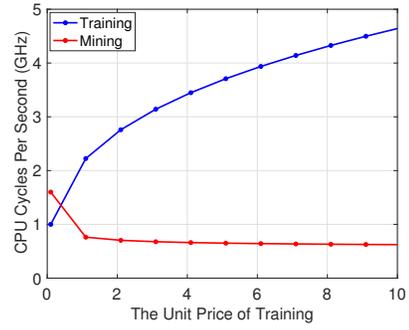}
\caption{CPU cycles per second for training and mining changing with $p_{ti}$.}
\label{fig:p_q}
\end{figure}

In the end, we explore the influence of $p_{ti}$ on both $q_{ti}$ and $q_{mi}$ to figure out  how the decisions of MO influence the decisions of client $i$. We set $\mu_i=10$ and  $p_{ti}\in [0, 10]$. In this setting, the simulation results are shown in Fig. \ref{fig:p_q}. We can see that the unit CPU cycles used in local model training has a positive relationship with the unit price of training offered by the MO, because more unit rewards for training will incentivize clients to put more computational power on model training. As for CPU cycles per second used in mining, it decreases with the increase of $p_{ti}$. This makes sense because if  clients are motivated to put more computing power into training,  the  training time will be reduced and the mining time will be correspondingly increased. In this way, clients do not need to set a high $q_{mi}$ for mining.

\section{Related Work}\label{sec:related}

Most of the existing studies related to BCFL focus on protecting privacy, achieving decentralization and improving the performance of model training \cite{peng2021vfchain,desai2021blockfla, lu2020blockchain, kim2018device,hu2021blockchain}. In our paper, we mainly focus on the resource allocation and incentive mechanism design in BCFL. Thus, we provide the literature review about these two areas in this section.

\subsection{Resource Allocation in BCFL}
As for resource allocation, researchers mainly consider the homogeneous computational power of all clients, and make decisions through the reinforcement learning approach. 

In \cite{Li2021}, the resource allocation problem is resolved for the local devices with the same computational power in BCFL. An upper-bound of the global loss function was proposed to evaluate the performance of training; in the meantime, the relationship among update rounds, block generation rate, and learning rate was explored. Although the proposed method can easily control the training and mining time by adjusting the number of updates to allocate resources, it is based on the assumption that all clients have the same amount of computing resources and local data, which is not practical. 

Hieu et al. \cite{Hieu2020} designed a deep reinforcement learning approach to help mobile devices determine the data volume and energy used for training and to assist the system to decide the block generation rate. 
Neither of the two works considers how to motivate clients to work honestly and efficiently.

According to the above discussion, it can be seen that the studies related to resource allocation in BCFL are are insufficient. One of the reasons is that the research regarding BCFL is still in the early stage. Another reason is that there are many types of BCFL structures depending on the role of the blockchain playing in FL, making it difficult to have a common framework for resource allocation. 
In order to assist the MO and the clients of the BCFL system to make the proper decisions, we design the mechanisms based on the two-stage Stackelberg game in this paper.
Besides, we consider allocating resources in the fully coupled BCFL with FL clients working as blockchain nodes. 

\subsection{Incentive Mechanism in BCFL}

There are some studies about BCFL that focus on regulating the behaviors of clients through incentive mechanism design, thus encouraging them to work honestly and efficiently according to the predefined rules.

Toyoda et al. \cite{toyoda2019mechanism} proposed an economic approach based on the assumption that clients would act rationally, where the repeated competition method was utilized to ensure that clients will follow the protocol. Bao et al. \cite{bao2019flchain} designed an incentive mechanism to attract more data and computational power contributing to the framework of BCFL. In their proposed system, honest clients can gain fairly partitioned rewards while the malicious clients will be punished via timely behaviour detection scheme. In \cite{Kang2019}, an incentive mechanism that integrated reputation and contract theory was proposed to encourage clients to provide high-quality data to train the local models. 
As for the fairness of reward allocation,  Liu et al. \cite{Liu2020} used Shapley Value (SV) to calculate the contributions of clients of the FL system and then allocate the rewards accordingly. 
However, this approach is not able to make incentive decisions for training and mining, respectively.


The existing studies about incentive mechanism design in BCFL focus on how to provide incentives for FL through blockchain, without considering the incentives for blockchain and FL in a systematical manner. In other words, blockchain and FL are in different phases for BCFL, so they should both have reasonable incentives. In our paper, we design a pricing mechanism for the MO based on the computing power provided by clients, thus providing incentives to the whole BCFL system.

In general, the existing studies have paid little attention to resource allocation for mobile devices in BCFL and assume that clients join the task voluntarily. To address this challenge, we design a resource allocation mechanism for mobile devices, which also offers reward suggestions to the MO so as to motivate clients to participate in BCFL.

\section{Conclusion}\label{sec:conclusion}
This paper studies the resource allocation of clients in BCFL by designing incentive mechanism. We describe the interactions between clients and the MO as a two-stage Stakelberg game. Within our model, clients with varying computing power can determine the resources to invest in training and mining based on the rewards provided by the MO through maximizing their utilities, while the MO can also obtain the optimal utility. Since the local training related information of clients may be not known to the MO, we further study the game model and optimal solutions in the incomplete information case. Numerous experimental results show that our proposed mechanisms are effective.


\begin{appendices}
\renewcommand{\thesection}{\Alph{section}.}
\section{Proof of Theorem \ref{eq:df2}} \label{ap_a}
The Lagrangian correlated to \textbf{Problem 1} is expressed as:
\begin{align}
\begin{aligned}
    \mathcal{L}_1
    &=\frac{\mu_i}{q_{ti}} p_{ti}+\frac{\psi}{q_{mi}} p_{mi}-\rho_i \mu_i q_{ti}^2-\rho_i \psi\Big (q_{mi}\Big)^2\\
    &-\lambda_1\Big(\rho_i \mu_i q_{ti}^2-\frac{\mu_i}{q_{ti}} p_{ti}\Big)-\lambda_2\Big(\rho_i \psi q_{mi}^2-\frac{\psi}{q_{mi}}p_{mi}\Big)\\
    &-\lambda_3\Big(\frac{\mu_i}{q_{ti}}+\frac{\psi}{q_{mi}}-T\Big), \forall i,\\
\end{aligned}
\end{align}
where $\lambda_1$, $\lambda_2$, and $\lambda_3$ are non-negative parameters correlated to the constraints of \textbf{Problem 1}.

The KKT conditions are as below:
\begin{align}
    \frac{\partial \mathcal{L}_1}{\partial q_{ti}} =\frac{\partial \mathcal{L}_1}{\partial q_{mi}}  & =0, \forall i,
    \label{eq:df_qt} \\
    \lambda_1 \geq 0, \lambda_2 \geq 0, \lambda_3  & \geq 0, \forall i,
    \label{eq:abc} \\
    \lambda_1\left(\rho_i \mu_i q_{ti}^2-\frac{\mu_i}{q_{ti}} p_{ti}\right)  & = 0,\forall i,
    \label{eq:lambda} \\
    \lambda_2\left(\rho_i \psi q_{mi}^2-\frac{\psi}{q_{mi}} p_{mi}\right) & =0, \forall i,
    \label{eq:alpha} \\
    \lambda_3\left(\frac{\mu_i}{q_{ti}}+\frac{\psi}{q_{mi}}-T\right) & =0, \forall i,
    \label{eq:lambda_3} \\
    \frac{\mu_i}{q_{ti}}^t p_{ti}-\rho_i \mu_i q_{ti}^2 & \geq 0, \forall i,
    \label{eq:reward_qt} \\
    \frac{\psi}{q_{mi}} p_{mi}-\rho_i \psi q_{mi}^2 & \geq 0, \forall i,
    \label{eq:reward_qm} \\
    \frac{\mu_i}{q_{ti}}+\frac{\psi}{q_{mi}} & \leq T, \forall i.
    \label{eq:time}
\end{align}
According to (\ref{eq:df_qt}), we can have $\frac{\partial \mathcal{L}_1}{\partial q_{ti}} =\frac{u_i\lambda_3}{q_{ti}^2}-2(1+\lambda_1)\rho_i u_i q_{ti}, \forall i$. Let the above equation equal to $0$ and we have
\begin{align}
    \frac{\lambda_3}{q_{ti}^2}=2(1+\lambda_1)\rho_i q_{ti}, \forall i.
    \label{eq:df_qt1}
\end{align}
Similarly, we have
\begin{align}
    \frac{\lambda_3}{q_{mi}^2}=2(1+\lambda_2)\rho_i q_{mi}, \forall i.
    \label{eq:df_qm1}
\end{align}
Then, let's consider equation (\ref{eq:time}). Assume that $\frac{\mu_i}{q_{ti}}+\frac{\psi}{q_{mi}}-T\neq 0, \forall i $, according to (\ref{eq:time}), we have $\lambda_3 = 0$. From (\ref{eq:df_qt1}), we can see that if $\lambda_3 = 0$, this equation will be $2(1+\lambda_1)\rho_i q_{ti}=0, \forall i$, then we have $\lambda_1=-1<0$. Since $\lambda_1$ is constrained by (\ref{eq:abc}), it should always be non-negative. Therefore, this assumption is invalid. We can obtain the same conclusion from (\ref{eq:df_qm1}) as well. So we can 
conclude that for any $i$, equation $\frac{\mu_i}{q_{ti}}+\frac{\psi}{q_{mi}}-T=0$ is always satisfied. In this way, $\lambda_3 > 0$ can be deduced.

Based on the KKT conditions and $\frac{\mu_i}{q_{ti}}+\frac{\psi}{q_{mi}}-T\neq 0, \forall i $, we can analyze the optimal solutions of Problem 1 as follows:

\textbf{Case 1}: $\lambda_1=\lambda_2= 0$, $\frac{\mu_i}{q_{ti}}+\frac{\psi}{q_{mi}}-T= 0, \forall i $.

In this case, since $\lambda_1=\lambda_2=0$, we can derive $q_{ti}=q_{mi}=\sqrt[3]{\frac{\lambda_3}{2\rho_i}}\geq 0$ using (\ref{eq:df_qt1}) and (\ref{eq:df_qm1}), respectively. But $\lambda_3$ is a non-negative parameter, and it is not a constant value, so we still can not get the optimal solutions of \textbf{Problem 1}. Thus, this case is not suitable.

\textbf{Case 2}: $\rho_i \mu_i q_{ti}^2-\frac{\mu_i}{q_{ti}} p_{ti}=\rho_i \psi q_{mi}^2-\frac{\psi}{q_{mi}} p_{mi}=0, \forall i$, $\frac{\mu_i}{q_{ti}}+\frac{\psi}{q_{mi}}-T= 0, \forall i $.

By solving $\rho_i \mu_i q_{ti}^2-\frac{\mu_i}{q_{ti}} p_{ti}=0, \forall i$ and $\rho_i \psi q_{mi}^2-\frac{\psi}{q_{mi}} p_{mi}=0, \forall i$, we have $q_{ti}=\sqrt[3]{\frac{p_{ti}}{\rho_i}},\forall i$ and $q_{mi}=\sqrt[3]{\frac{p_{mi}}{\rho_i}},\forall i$. 


Since $\frac{\mu_i}{q_{ti}}+\frac{\psi}{q_{mi}}-T=0, \forall i$, even though the above two functions can give the expression of the solution of \textbf{Problem 1}, it is still constrained by this function. In other words, the one of the KKT conditions, i.e., (\ref{eq:time}), is not satisfied. Thus, this case is not suitable.
 
\textbf{Case 3}: $\left(\rho_i \mu_i q_{ti}^2-\frac{\mu_i}{q_{ti}} p_{ti}\right) = 0$, $\lambda_2=0$, $\frac{\mu_i}{q_{ti}}+\frac{\psi}{q_{mi}}-T=0, \forall i$.

From (\ref{eq:time}), we can get the relationship between $q_{ti}$ and $q_{mi}$ is $q_{mi}=\frac{\psi}{T-\frac{\mu_i}{q_{ti}}}$. Solving $\left(\rho_i \mu_i q_{ti}^2-\frac{\mu_i}{q_{ti}} p_{ti}\right) = 0$ yields $q_{ti} = \sqrt[3]{\frac{p_{ti}}{\rho_i}}, \forall i$. Based on $q_{mi}=\frac{\psi}{T-\frac{\mu_i}{q_{ti}}}$, we let $q_{ti} = \sqrt[3]{\frac{p_{ti}}{\rho_i}}, \forall i$, then we can derive that $q_m(t)=\frac{\psi}{T-\frac{\mu_i}{\sqrt[3]{\frac{p_{ti}}{\rho_i}}}}, \forall i.$ From (\ref{eq:df_qt1}) and (\ref{eq:df_qm1}), we have $\lambda_1=\frac{\lambda_3}{2\rho_i q_{ti}^3}-1$ and $\lambda_2=\frac{\lambda_3}{2\rho_i q_{mi}^3}-1$. Since $\lambda_3$, $q_{ti}$, $q_{mi}$ and $\rho_i$ are positive, so $\lambda_1=\frac{\lambda_3}{2\rho_i q_{ti}^3}>0$, and $\lambda_3$ can be large enough to make sure $\lambda_1=\frac{\lambda_3}{2\rho_i q_{ti}^3}\geq 1$, thus $\lambda_1\geq0$ can be guaranteed. Similarly, $\lambda_2\geq0$ can be derived. From the above analysis, \textbf{Case 3} satisfies all the KKT conditions, therefore the optimal solutions are obtained.

\textbf{Case 4}: $\left(\rho_i \psi q_{mi}^2-\frac{\psi}{q_{mi}} p_{mi}\right)=0, \forall i$, $\lambda_1=0$, $\frac{\mu_i}{q_{ti}}+\frac{\psi}{q_{mi}}-T=0, \forall i$. This case is similar to \textbf{Case 3}.

Based on the above analysis, the optimal solutions of \textbf{Problem 1} are $q_{ti}^* = \left({\frac{p_{ti}}{\rho_i}}\right)^\frac{1}{3}, \forall i,$ and $q_{mi}^*=\frac{\psi}{T- \mu_i \left(\frac{\rho_i}{p_{ti}}\right)^\frac{1}{3}}$. 

Thus \textbf{Theorem \ref{eq:df2}} is proved.

\section{Proof of Theorem \ref{def 2}}
The Lagrangian correlated to Problem 2 is
\begin{align}
\begin{aligned}
    \mathcal{L}_2
    &= f\left(\sum_{i=1}^{N} \mu_i\right)-\xi \sum_{i=1}^{N} \left(\frac{\mu_i}{q_{ti}}+\frac{\psi}{q_{mi}}+\frac{\mu_i}{q_{ti}} p_{ti}+\frac{\psi}{q_{mi}} p_{mi}\right)\\
    &-\theta_1 \left(\frac{\mu_i}{q_{ti}}p_ti-\overline{R}_{ti}\right)-\theta_2 \left(\frac{\psi}{q_mi}p_mi-\overline{R}_{mi}\right), 
\label{eq:l_mo}
\end{aligned}
\end{align}
where $\theta_1$ and $\theta_2$ are the Lagrange multipliers correlated to the constraints of $\mathrm{Problem 2}$. The following constrains should be met:
\begin{align}
  \frac{\partial \mathcal{L}_2}{\partial p_{ti}} =\frac{\partial \mathcal{L}_2}{\partial p_{mi}}  & =0, \forall i,
  \label{eq:df_pt} \\
    \theta_1 \geq 0, \theta_2  & \geq 0,\forall i, \\
    \theta_1 \left(\frac{\mu_i}{q_{ti}}p_{ti}-\overline{R}_{ti}\right) & =0,\forall i, \\
    \theta \left(\frac{\mu_i}{q_{ti}} p_{ti}+\frac{\psi}{q_{mi}} p_{mi}-\omega\right) & =0, \forall i, \\
    \frac{\mu_i}{q_{ti}} p_{ti}+\frac{\psi}{q_{mi}} p_{mi} & \leq \omega, \forall i. \\
    \frac{\mu_i}{q_{ti}}p_ti  & \leq \overline{R}_{ti},\forall i,
    \label{eq:r_t} \\
    \frac{\psi}{q_mi}p_{mi} & \leq \overline{R}_{mi}, \forall i.
    \label{eq:r_m}
\end{align}

First, let $q_{ti}=q_{ti}^*$ and $q_{mi}=q_i(n)^*$.

\textbf{Case 1:} $\theta_1=0$, $\frac{\psi}{q_mi}p_{mi}-\overline{R}_{mi}=0, \forall i$.

In this case, we can have $\frac{\partial \mathcal{L}_2}{\partial p_{ti}} =\frac{-(\mu_i(\xi p_{mi}+2\xi p_{ti}+\theta_2 p_{mi}))}{3p_{ti}\sqrt[3]{\frac{p_{ti}}{\rho_i} } } , \forall i$. Setting this equation equal to $0$ yields $p_{ti}=\frac{-p_{mi}(\xi + \theta_2)}{2\xi}$. Obviously, we cannot find a positive $\theta_2$ to satisfy this equation, making this solution invalid.

\textbf{Case 2:} $\theta_2=0, \frac{\psi}{q_{ti}}p_{ti}-\overline{R}_{ti}=0, \forall i$. 

This case is similar to Case 1.

Case 3: $\frac{\psi}{q_{ti}}p_{ti}-\overline{R}_{ti}=0$,  $\frac{\psi}{q_mi}p_{mi}-\overline{R}_{mi}=0, \forall i$. 

By solving  $\frac{\psi}{q_{ti}}p_{ti}-\overline{R}_{ti}=0$,  $\frac{\psi}{q_mi}p_{mi}-\overline{R}_{mi}=0, \forall i$, we can get (\ref{eq:optimal_p_t}) and (\ref{eq:optimal_p_m}).  We can also prove that this case satisfy the rest of the KKT conditions. 

Thus, Theorem \ref{def 2} is proved.

\section{Proof of theorem \ref{th_3}}
Then, we will provide the solution of Problem 3. The Lagrangian of Problem 3 can be written as
\begin{align}
    \mathcal{L}_3
    &=f \left(\sum_{i=1}^{N} \mu_i\right)-\xi \sum_{i=1}^{N} \left(\frac{\mu_i}{q_{ti}}+\frac{\psi}{q_{mi}}+\frac{\mu_i}{q_{ti}} p_{ti}+\frac{\psi}{q_{mi}} p_{mi}\right)\notag\\
    &-\alpha_1 \left(\frac{\mu_i}{q_{ti}}p_ti-\overline{R}_{ti}\right)-\alpha_2 \left(\frac{\psi}{q_mi}p_mi-\overline{R}_{mi}\right)\notag\\
    &-\alpha_3 \left(\frac{\mu_i}{q_{ti}}p_{ti}+\frac{\psi}{q_{mi}} p_{mi}-\rho_i \mu_i q_{ti}^2-\rho_i \psi q_{mi}^2\right).
\label{eq:l_mo_3}
\end{align}
where $\alpha_1$, $\alpha_2$ and $\alpha_3$ are the Lagrange multipliers.
The KKT conditions are similar with Problem 2 except the following three conditions:
\begin{align}
\alpha_3  & \geq 0, \forall i, \\
\alpha_3 \left(\frac{\mu_i}{q_{ti}}p_{ti}+\frac{\psi}{q_{mi}} p_{mi}-\rho_i \mu_i q_{ti}^2-\rho_i \psi q_{mi}^2\right) & =0, \forall i, \\
\frac{\mu_i}{q_{ti}}p_{ti}+\frac{\psi}{q_{mi}} p_{mi}-\rho_i \mu_i q_{ti}^2-\rho_i \psi q_{mi}^2 & \geq 0, \forall i.
\end{align}

We then analyze the solutions under different cases.
Actually, there should be nine cases in this problem, but we only give two of them to analyze since the other situation can be interpreted similarly.

$\mathbf{Case 1:}$ $\alpha_1 = \alpha_2=\alpha_3 = 0, \forall i.$

In this case, we can have $ \frac{\partial \mathcal{L}_3}{\partial p_{ti}} =\frac{-(\mu_i\xi(p_{mi}+2p_{ti}))}{3\rho_i\sqrt[\frac{4}{3}]{\frac{p_{ti}}{\rho_i}}}$, and let it equal to $0$ we can get $p_{ti} = \frac{-p_{mi}}{2}$. Obviously, since $p_{ti}$ and $p_{mi}$ are non-negative values,  we cannot find a $p_{mi}$ to satisfy the above equation. So this case is invalid.

$\mathbf{Case 2:}$ $\frac{\mu_i}{q_{ti}}p_{ti}+\frac{\psi}{q_{mi}} p_{mi}-\rho_i \mu_i q_{ti}^2-\rho_i \psi q_{mi}^2=0$, $\frac{\psi}{q_{ti}}p_{ti}-\overline{R}_{ti}=0$,  $\alpha_2=0, \forall i.$
By solving the above equations, we get  (\ref{eq:optimal_p_t_2}) and (\ref{eq:optimal_p_m_2}).
We can verify that the solutions above is incentive compatible and satisfy all the KKT conditions. 

Thus Theorem \ref{th_3} is proved.

\end{appendices}
\bibliographystyle{IEEEtran}
\bibliography{reference}

\begin{thebibliography}{10}
\providecommand{\url}[1]{#1}
\csname url@samestyle\endcsname
\providecommand{\newblock}{\relax}
\providecommand{\bibinfo}[2]{#2}
\providecommand{\BIBentrySTDinterwordspacing}{\spaceskip=0pt\relax}
\providecommand{\BIBentryALTinterwordstretchfactor}{4}
\providecommand{\BIBentryALTinterwordspacing}{\spaceskip=\fontdimen2\font plus
\BIBentryALTinterwordstretchfactor\fontdimen3\font minus
  \fontdimen4\font\relax}
\providecommand{\BIBforeignlanguage}[2]{{%
\expandafter\ifx\csname l@#1\endcsname\relax
\typeout{** WARNING: IEEEtran.bst: No hyphenation pattern has been}%
\typeout{** loaded for the language `#1'. Using the pattern for}%
\typeout{** the default language instead.}%
\else
\language=\csname l@#1\endcsname
\fi
#2}}
\providecommand{\BIBdecl}{\relax}
\BIBdecl

\bibitem{zhao2020mobile}
Y.~Zhao, J.~Zhao, L.~Jiang, R.~Tan, and D.~Niyato, ``Mobile edge computing,
  blockchain and reputation-based crowdsourcing iot federated learning: A
  secure, decentralized and privacy-preserving system,'' 2020.

\bibitem{zhang2020blockchain}
W.~Zhang, Q.~Lu, Q.~Yu, Z.~Li, Y.~Liu, S.~K. Lo, S.~Chen, X.~Xu, and L.~Zhu,
  ``Blockchain-based federated learning for device failure detection in
  industrial iot,'' \emph{IEEE Internet of Things Journal}, vol.~8, no.~7, pp.
  5926--5937, 2020.

\bibitem{lu2019blockchain}
Y.~Lu, X.~Huang, Y.~Dai, S.~Maharjan, and Y.~Zhang, ``Blockchain and federated
  learning for privacy-preserved data sharing in industrial iot,'' \emph{IEEE
  Transactions on Industrial Informatics}, vol.~16, no.~6, pp. 4177--4186,
  2019.

\bibitem{qi2021privacy}
Y.~Qi, M.~S. Hossain, J.~Nie, and X.~Li, ``Privacy-preserving blockchain-based
  federated learning for traffic flow prediction,'' \emph{Future Generation
  Computer Systems}, vol. 117, pp. 328--337, 2021.

\bibitem{hua2020blockchain}
G.~Hua, L.~Zhu, J.~Wu, C.~Shen, L.~Zhou, and Q.~Lin, ``Blockchain-based
  federated learning for intelligent control in heavy haul railway,''
  \emph{IEEE Access}, vol.~8, pp. 176\,830--176\,839, 2020.

\bibitem{passerat2019blockchain}
J.~Passerat-Palmbach, T.~Farnan, R.~Miller, M.~S. Gross, H.~L. Flannery, and
  B.~Gleim, ``A blockchain-orchestrated federated learning architecture for
  healthcare consortia,'' \emph{arXiv preprint arXiv:1910.12603}, 2019.

\bibitem{aich2021protecting}
S.~Aich, N.~K. Sinai, S.~Kumar, M.~Ali, Y.~R. Choi, M.-I. Joo, and H.-C. Kim,
  ``Protecting personal healthcare record using blockchain \& federated
  learning technologies,'' in \emph{2021 23rd International Conference on
  Advanced Communication Technology (ICACT)}.\hskip 1em plus 0.5em minus
  0.4em\relax IEEE, 2021, pp. 109--112.

\bibitem{kumar2021blockchain}
R.~Kumar, A.~A. Khan, J.~Kumar, A.~Zakria, N.~A. Golilarz, S.~Zhang, Y.~Ting,
  C.~Zheng, and W.~Wang, ``Blockchain-federated-learning and deep learning
  models for covid-19 detection using ct imaging,'' \emph{IEEE Sensors
  Journal}, 2021.

\bibitem{mcmahan2017communication}
B.~McMahan, E.~Moore, D.~Ramage, S.~Hampson, and B.~A. y~Arcas,
  ``Communication-efficient learning of deep networks from decentralized
  data,'' in \emph{Artificial intelligence and statistics}.\hskip 1em plus
  0.5em minus 0.4em\relax PMLR, 2017, pp. 1273--1282.

\bibitem{bagdasaryan2020backdoor}
E.~Bagdasaryan, A.~Veit, Y.~Hua, D.~Estrin, and V.~Shmatikov, ``How to backdoor
  federated learning,'' in \emph{International Conference on Artificial
  Intelligence and Statistics}.\hskip 1em plus 0.5em minus 0.4em\relax PMLR,
  2020, pp. 2938--2948.

\bibitem{sattler2019robust}
F.~Sattler, S.~Wiedemann, K.-R. M{\"u}ller, and W.~Samek, ``Robust and
  communication-efficient federated learning from non-iid data,'' \emph{IEEE
  transactions on neural networks and learning systems}, vol.~31, no.~9, pp.
  3400--3413, 2019.

\bibitem{lyu2020threats}
L.~Lyu, H.~Yu, and Q.~Yang, ``Threats to federated learning: A survey,''
  \emph{arXiv preprint arXiv:2003.02133}, 2020.

\bibitem{mothukuri2021survey}
V.~Mothukuri, R.~M. Parizi, S.~Pouriyeh, Y.~Huang, A.~Dehghantanha, and
  G.~Srivastava, ``A survey on security and privacy of federated learning,''
  \emph{Future Generation Computer Systems}, vol. 115, pp. 619--640, 2021.

\bibitem{ramanan2020baffle}
P.~Ramanan and K.~Nakayama, ``Baffle: Blockchain based aggregator free
  federated learning,'' in \emph{2020 IEEE International Conference on
  Blockchain (Blockchain)}.\hskip 1em plus 0.5em minus 0.4em\relax IEEE, 2020,
  pp. 72--81.

\bibitem{kim2019blockchain}
Y.~J. Kim and C.~S. Hong, ``Blockchain-based node-aware dynamic weighting
  methods for improving federated learning performance,'' in \emph{2019 20th
  Asia-Pacific Network Operations and Management Symposium (APNOMS)}.\hskip 1em
  plus 0.5em minus 0.4em\relax IEEE, 2019, pp. 1--4.

\bibitem{liu2020fedcoin}
Y.~Liu, Z.~Ai, S.~Sun, S.~Zhang, Z.~Liu, and H.~Yu, ``Fedcoin: A peer-to-peer
  payment system for federated learning,'' in \emph{Federated Learning}.\hskip
  1em plus 0.5em minus 0.4em\relax Springer, 2020, pp. 125--138.

\bibitem{Li2021}
\BIBentryALTinterwordspacing
J.~Li, Y.~Shao, K.~Wei, M.~Ding, C.~Ma, L.~Shi, Z.~Han, and H.~V. Poor,
  ``{Blockchain Assisted Decentralized Federated Learning (BLADE-FL):
  Performance Analysis and Resource Allocation},'' pp. 1--12, 2021. [Online].
  Available: \url{http://arxiv.org/abs/2101.06905}
\BIBentrySTDinterwordspacing

\bibitem{Hieu2020}
\BIBentryALTinterwordspacing
N.~Q. Hieu, T.~T. Anh, N.~C. Luong, D.~Niyato, D.~I. Kim, and E.~Elmroth,
  ``{Resource Management for Blockchain-enabled Federated Learning: A Deep
  Reinforcement Learning Approach},'' 2020. [Online]. Available:
  \url{http://arxiv.org/abs/2004.04104}
\BIBentrySTDinterwordspacing

\bibitem{qu2020privacy}
X.~Qu, Q.~Hu, and S.~Wang, ``Privacy-preserving model training architecture for
  intelligent edge computing,'' \emph{Computer Communications}, vol. 162, pp.
  94--101, 2020.

\bibitem{wang2021blockchain}
Z.~Wang and Q.~Hu, ``Blockchain-based federated learning: A comprehensive
  survey,'' \emph{arXiv preprint arXiv:2110.02182}, 2021.

\bibitem{nakamoto2008bitcoin}
S.~Nakamoto, ``Bitcoin: A peer-to-peer electronic cash system,''
  \emph{Decentralized Business Review}, p. 21260, 2008.

\bibitem{castro1999practical}
M.~Castro, B.~Liskov \emph{et~al.}, ``Practical byzantine fault tolerance,'' in
  \emph{OSDI}, vol.~99, no. 1999, 1999, pp. 173--186.

\bibitem{mingxiao2017review}
D.~Mingxiao, M.~Xiaofeng, Z.~Zhe, W.~Xiangwei, and C.~Qijun, ``A review on
  consensus algorithm of blockchain,'' in \emph{2017 IEEE international
  conference on systems, man, and cybernetics (SMC)}.\hskip 1em plus 0.5em
  minus 0.4em\relax IEEE, 2017, pp. 2567--2572.

\bibitem{burd1996processor}
T.~D. Burd and R.~W. Brodersen, ``Processor design for portable systems,''
  \emph{Journal of VLSI signal processing systems for signal, image and video
  technology}, vol.~13, no.~2, pp. 203--221, 1996.

\bibitem{zhang2009stackelberg}
J.~Zhang and Q.~Zhang, ``Stackelberg game for utility-based cooperative
  cognitiveradio networks,'' in \emph{Proceedings of the tenth ACM
  international symposium on Mobile ad hoc networking and computing}, 2009, pp.
  23--32.

\bibitem{peng2021vfchain}
Z.~Peng, J.~Xu, X.~Chu, S.~Gao, Y.~Yao, R.~Gu, and Y.~Tang, ``Vfchain: Enabling
  verifiable and auditable federated learning via blockchain systems,''
  \emph{IEEE Transactions on Network Science and Engineering}, 2021.

\bibitem{desai2021blockfla}
H.~B. Desai, M.~S. Ozdayi, and M.~Kantarcioglu, ``Blockfla: Accountable
  federated learning via hybrid blockchain architecture,'' in \emph{Proceedings
  of the Eleventh ACM Conference on Data and Application Security and Privacy},
  2021, pp. 101--112.

\bibitem{lu2020blockchain}
Y.~Lu, X.~Huang, K.~Zhang, S.~Maharjan, and Y.~Zhang, ``Blockchain empowered
  asynchronous federated learning for secure data sharing in internet of
  vehicles,'' \emph{IEEE Transactions on Vehicular Technology}, vol.~69, no.~4,
  pp. 4298--4311, 2020.

\bibitem{kim2018device}
H.~Kim, J.~Park, M.~Bennis, and S.-L. Kim, ``On-device federated learning via
  blockchain and its latency analysis,'' \emph{arXiv preprint
  arXiv:1808.03949}, 2018.

\bibitem{hu2021blockchain}
Q.~Hu, Z.~Wang, M.~Xu, and X.~Cheng, ``Blockchain and federated edge learning
  for privacy-preserving mobile crowdsensing,'' \emph{IEEE Internet of Things
  Journal}, 2021.

\bibitem{toyoda2019mechanism}
K.~Toyoda and A.~N. Zhang, ``Mechanism design for an incentive-aware
  blockchain-enabled federated learning platform,'' in \emph{2019 IEEE
  International Conference on Big Data (Big Data)}.\hskip 1em plus 0.5em minus
  0.4em\relax IEEE, 2019, pp. 395--403.

\bibitem{bao2019flchain}
X.~Bao, C.~Su, Y.~Xiong, W.~Huang, and Y.~Hu, ``Flchain: A blockchain for
  auditable federated learning with trust and incentive,'' in \emph{2019 5th
  International Conference on Big Data Computing and Communications
  (BIGCOM)}.\hskip 1em plus 0.5em minus 0.4em\relax IEEE, 2019, pp. 151--159.

\bibitem{Kang2019}
J.~Kang, Z.~Xiong, D.~Niyato, S.~Xie, and J.~Zhang, ``{Incentive Mechanism for
  Reliable Federated Learning: A Joint Optimization Approach to Combining
  Reputation and Contract Theory},'' \emph{IEEE Internet of Things Journal},
  vol.~6, no.~6, pp. 10\,700--10\,714, 2019.

\bibitem{Liu2020}
\BIBentryALTinterwordspacing
Y.~Liu, S.~Sun, Z.~Ai, S.~Zhang, Z.~Liu, and H.~Yu, ``{FedCoin: A Peer-to-Peer
  Payment System for Federated Learning},'' 2020. [Online]. Available:
  \url{http://arxiv.org/abs/2002.11711}
\BIBentrySTDinterwordspacing

\end{thebibliography}

\begin{IEEEbiography}[{\includegraphics[width=1in,height=1.25in,clip,keepaspectratio]{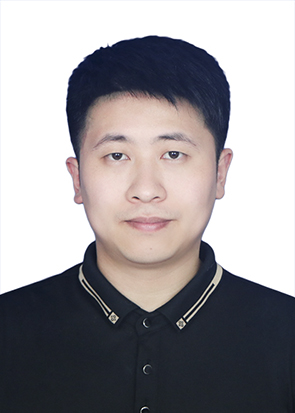}}]{Zhilin Wang} received his B.S. from Nanchang University in 2020. He is currently pursuing his Ph.D. degree of Computer and Information Science In Indiana University-Purdue University Indianapolis (IUPUI). He is a Research Assistant with IUPUI, and he is also a reviewer of 2022 IEEE International Conference on Communications (ICC). His research interests include blockchain, federated learning, edge computing, and Internet of Things (IoT).
\end{IEEEbiography}

\begin{IEEEbiography}[{\includegraphics[width=1in,height=1.25in,clip,keepaspectratio]{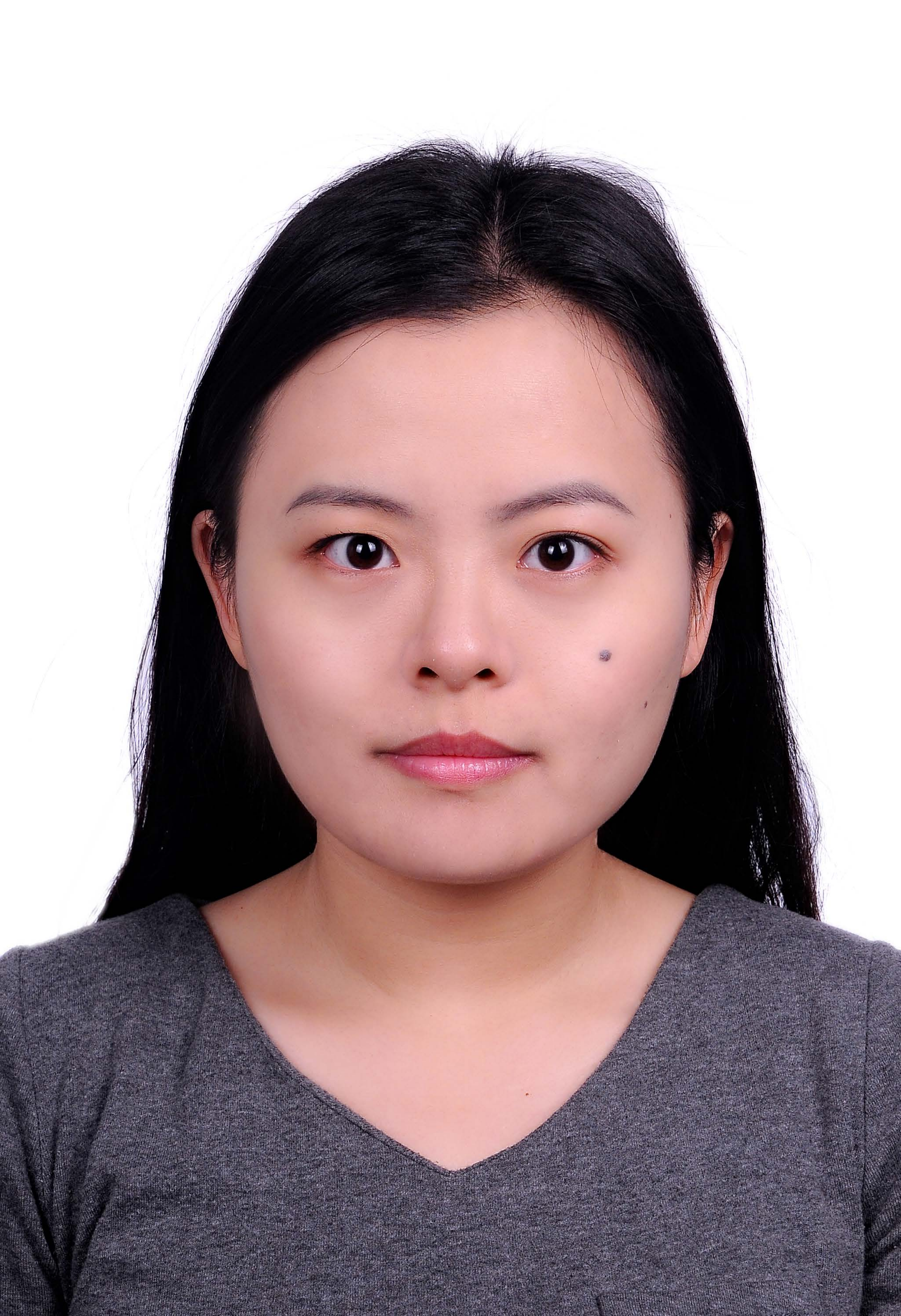}}]{Qin Hu} received her Ph.D. degree in Computer Science from the George Washington University in 2019. She is currently an Assistant Professor with the Department of Computer and Information Science, Indiana University-Purdue University Indianapolis (IUPUI). She has served on the Editorial Board of two journals, the Guest Editor for two journals, the TPC/Publicity Co-chair for several workshops, and the TPC Member for several international conferences. Her research interests include wireless and mobile security, edge computing, blockchain, and crowdsensing.
\end{IEEEbiography}

\begin{IEEEbiography}[{\includegraphics[width=1in,height=1.25in,clip,keepaspectratio]{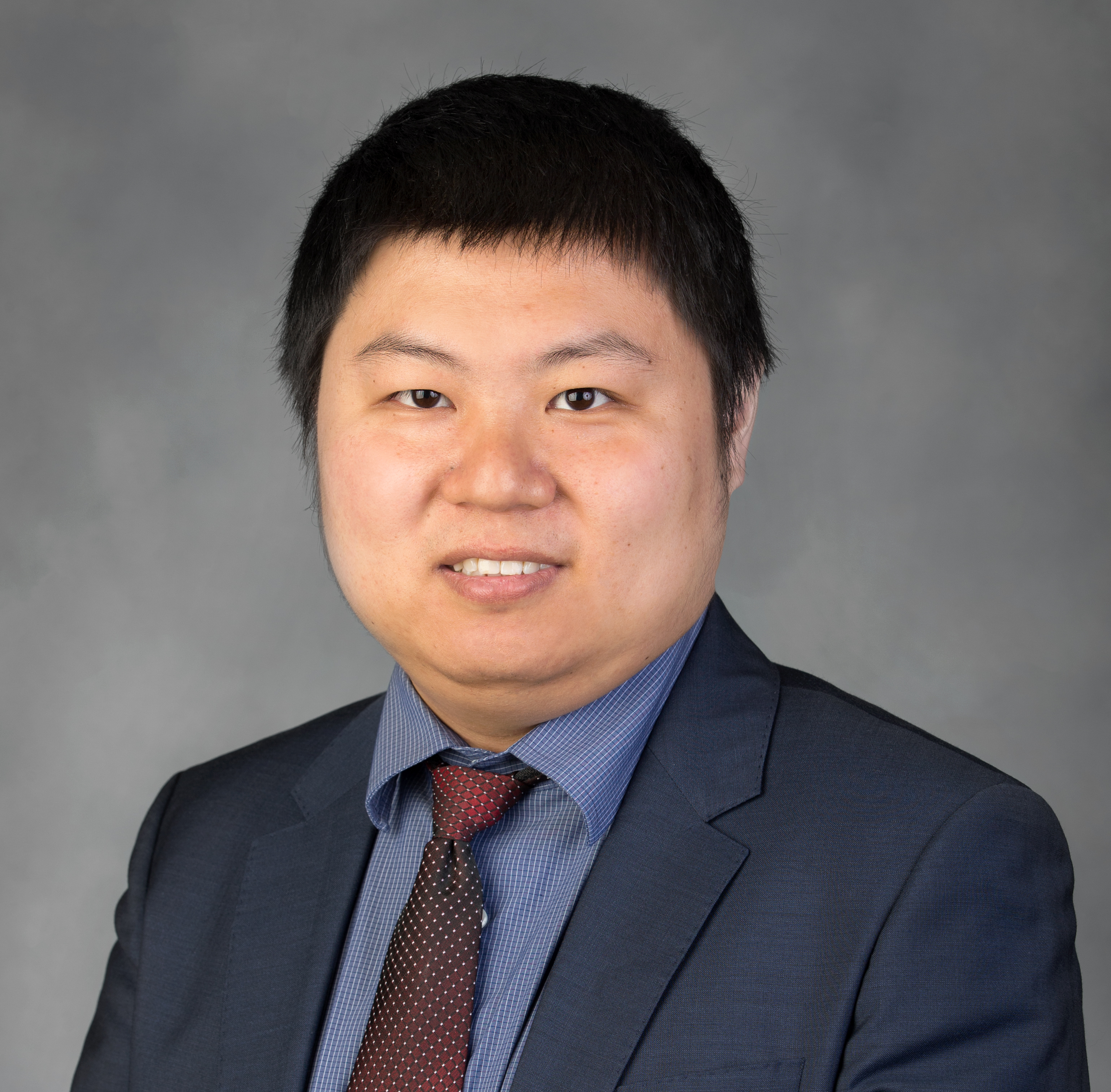}}]{Ruinian Li} received the PhD degree in computer
science from the George Washington University in
2018. He is currently an assistant professor at the
Department of Computer Science, Bowling Green
State University (BGSU), USA. His research interests include security and privacy-preserving computations, applied cryptography, and blockchain technology. He has been working in a wide area of
social networks, auction systems, and IoT, and his
work has been published in top-tier journals, such
as IEEE Transactions on Services Computing, and
IEEE Transactions on Network Science and Engineering.
\end{IEEEbiography}

\begin{IEEEbiography}[{\includegraphics[width=1in,height=1.25in,clip,keepaspectratio]{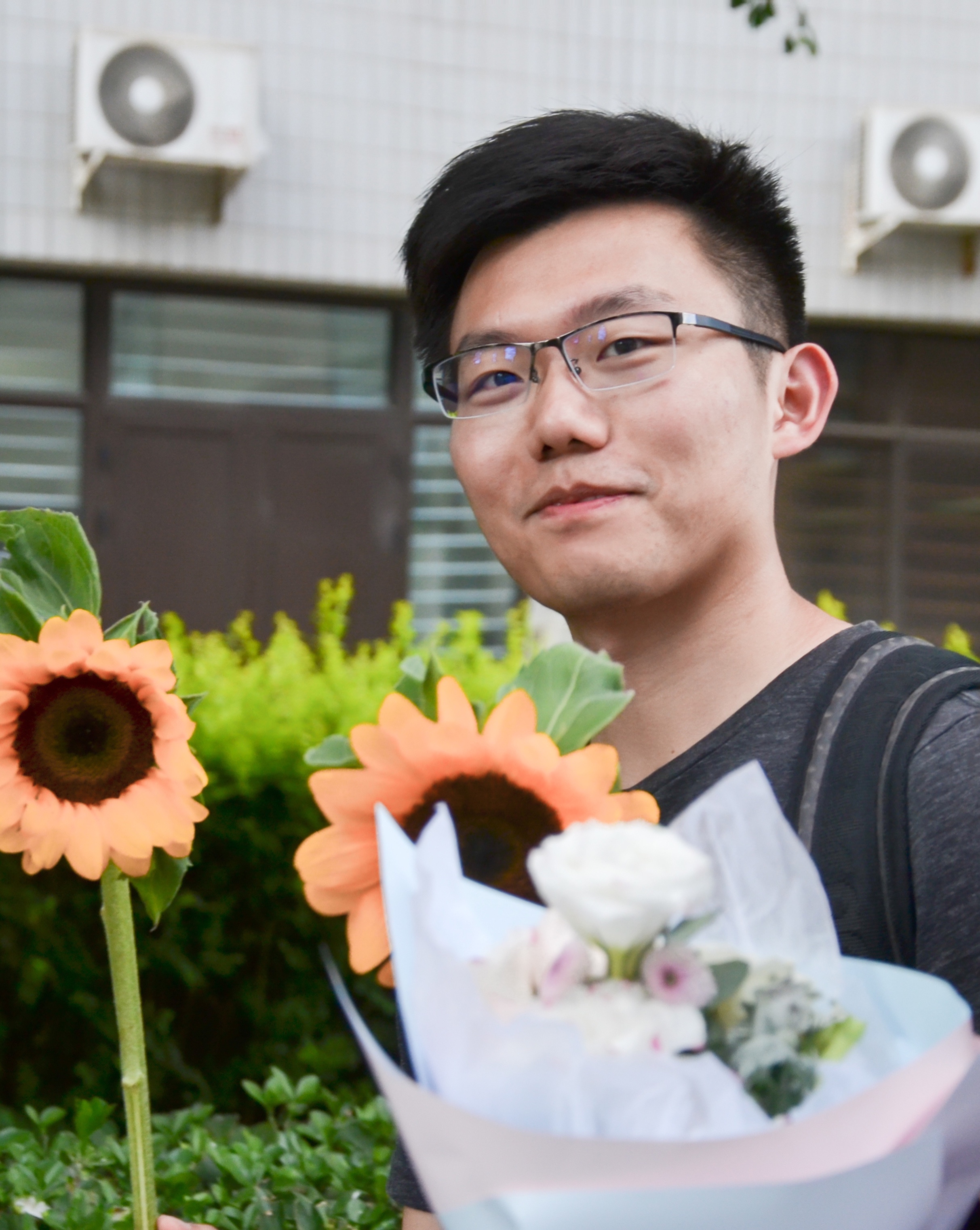}}]{Minghui Xu} received the BS degree in Physics from the Beijing Normal University, Beijing, China, in 2018, and the PhD degree in Computer Science from The George Washington University, Washington DC, USA, in 2021. He is currently an Assistant Professor in the School of Computer Science and Technology, Shandong University, China. His current research focuses on blockchain, distributed computing,  and quantum computing.
\end{IEEEbiography}

\begin{IEEEbiography}[{\includegraphics[width=1in,height=1.25in,clip,keepaspectratio]{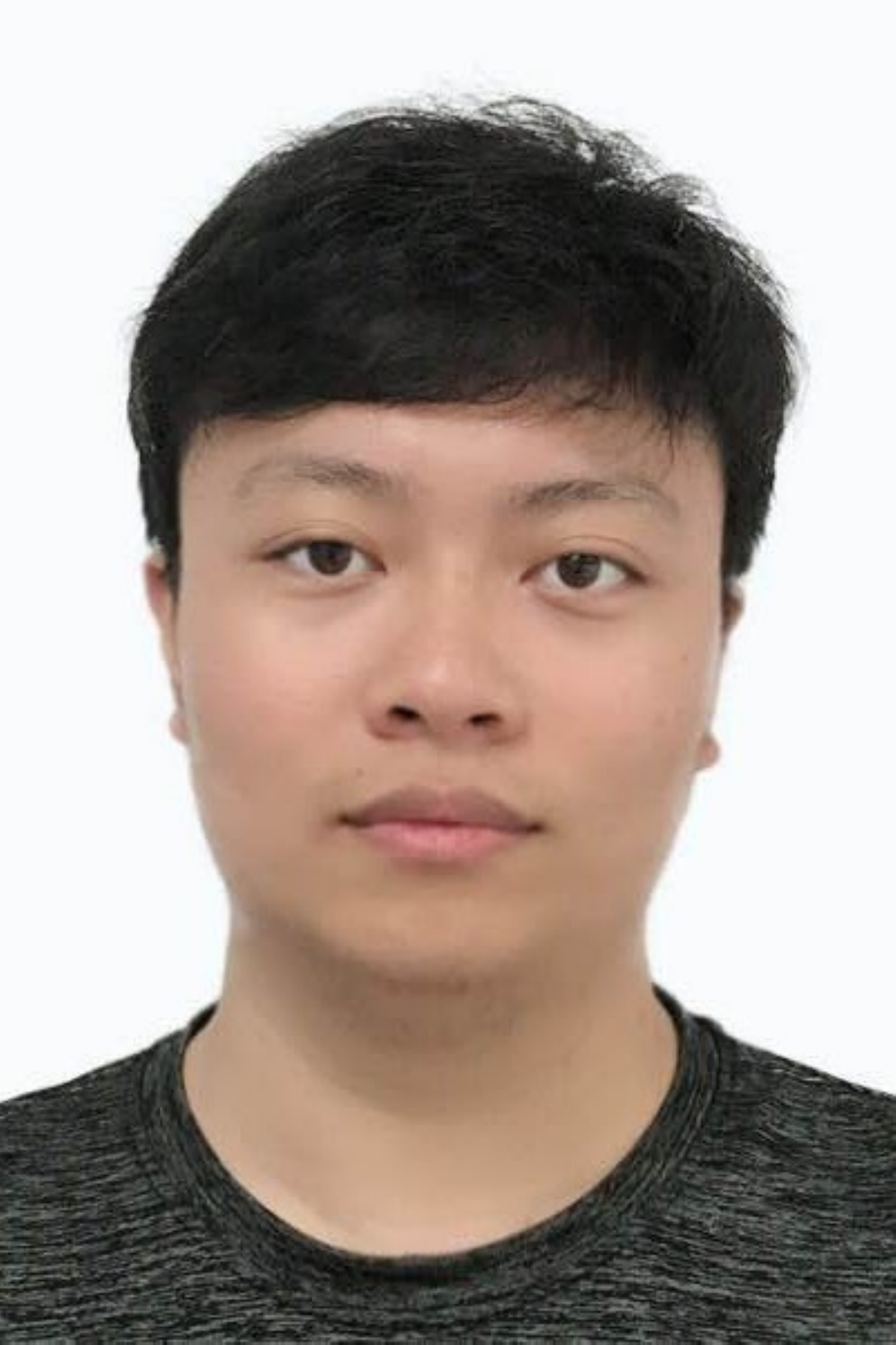}}]{Zehui Xiong} is currently an Assistant Professor in the Pillar of Information Systems Technology and Design, Singapore University of Technology and Design. Prior to that, he was a researcher with Alibaba-NTU Joint Research Institute, Singapore. He received the PhD degree in Nanyang Technological University, Singapore. He was the visiting scholar at Princeton University and University of Waterloo. His research interests include wireless communications, network games and economics, blockchain, and edge intelligence. He has published more than 140 research papers in leading journals and flagship conferences and many of them are ESI Highly Cited Papers. He has won over 10 Best Paper Awards in international conferences and is listed in the World’s Top $2\%$ Scientists identified by Stanford University. He is now serving as the editor or guest editor for many leading journals including IEEE JSAC, TVT, IoTJ, TCCN, TNSE, ISJ, JAS. He is the recipient of IEEE TCSC Early Career Researcher Award for Excellence in Scalable Computing, IEEE CSIM Technical Committee Best Journal Paper Award, IEEE SPCC Technical Committee Best Paper Award, IEEE VTS Singapore Best Paper Award, Chinese Government Award for Outstanding Students Abroad, and NTU SCSE Best PhD Thesis Runner-Up Award. He is the Founding Vice Chair of Special Interest Group on Wireless Blockchain Networks in IEEE Cognitive Networks Technical Committee.

\end{IEEEbiography}

\end{document}